\documentclass[reprint,superscriptaddress,amsmath,amssymb,aps,prb,showpacs]{revtex4-1}
\usepackage{float}
\usepackage[english]{babel}
\usepackage{multirow}
\usepackage[thinspace]{SIunits}
\usepackage{xspace}
\usepackage[normalem]{ulem} 
\usepackage{soul} 
\usepackage{graphicx}
\graphicspath{{figures//}} 
\usepackage{dcolumn}
\usepackage{bm}
\usepackage{color}
\usepackage{hyperref}
\usepackage{array}
\usepackage{booktabs}
\usepackage{multirow}
\hypersetup{
    unicode=false,          
    pdftoolbar=true,        
    pdfmenubar=true,        
    pdffitwindow=false,     
    pdfstartview={FitH},    
    pdftitle={My title},    
    pdfauthor={Author},     
    pdfsubject={Subject},   
    pdfcreator={Creator},   
    pdfproducer={Producer}, 
    pdfkeywords={keyword1} {key2} {key3}, 
    pdfnewwindow=true,      
    colorlinks=true,       
    linkcolor=blue,          
    citecolor=blue,        
    filecolor=blue,      
    urlcolor=blue,       
    plainpages=false        
}

\newcommand{\etal}{\textit{et al.}\xspace}
\newcommand{\red}[1]{\textcolor{red}{#1}}

\let\red\relax

\newcommand{\Tc}{$T_c$\xspace}

\newcommand{\BFCA}{\mbox{Ba(Fe$_{\mathrm{1-x}}$Co$_{\mathrm{x}}$)$_{2}$As$_2$}\xspace}

\newcommand{\BKFA}{\mbox{Ba$_{1-x}$K$_x$Fe$_{2}$As$_2$}\xspace}

\newcommand{\FeSelllll}{\mbox{(Li$_{1-x}$Fe$_{x}$)OHFeSe}\xspace}

\newcommand{\Alg}{${A_{1g}}$\xspace}

\newcommand{\Blg}{${B_{1g}}$\xspace}
\newcommand{\BZg}{${B_{2g}}$\xspace}

\newcommand{\Eg}{${E_{g}}$\xspace}

\newcommand{\wn}{$\,\mathrm{cm}^{-1}$\xspace}
\newcommand{\rev}[1]{\textcolor{red}{#1}}
\let\rev\relax

\begin{document}

\title{ \Large Raman Study of Cooper Pairing Instabilities in \FeSelllll}

\author{G. He}\email[Corresponding author: ]{Ge.He@wmi.badw.de}
\affiliation{Walther Meissner Institut, Bayerische Akademie der Wissenschaften, 85748 Garching, Germany}

\author{D. Li}
\affiliation{Beijing National Laboratory for Condensed Matter Physics, Institute of Physics, Chinese Academy of Sciences, Beijing 100190, China.}
\affiliation{School of Physical Sciences, University of Chinese Academy of Sciences, Beijing 100049, China.}

\author{D. Jost}
\affiliation{Walther Meissner Institut, Bayerische Akademie der Wissenschaften,
85748 Garching, Germany}
\affiliation{Fakult\"at f\"ur Physik E23, Technische Universit\"at M\"unchen,
85748 Garching, Germany}

\author{A. Baum}
\affiliation{Walther Meissner Institut, Bayerische Akademie der Wissenschaften,
85748 Garching, Germany}

\author{P. P. Shen}
\affiliation{Beijing National Laboratory for Condensed Matter Physics, Institute of Physics, Chinese Academy of Sciences, Beijing 100190, China.}
\affiliation{School of Physical Sciences, University of Chinese Academy of Sciences, Beijing 100049, China.}

\author{X. L. Dong}
\affiliation{Beijing National Laboratory for Condensed Matter Physics, Institute of Physics, Chinese Academy of Sciences, Beijing 100190, China.}
\affiliation{School of Physical Sciences, University of Chinese Academy of Sciences, Beijing 100049, China.}
\affiliation{Songshan Lake Materials Laboratory, Dongguan, Guangdong 523808, China}

\author{Z. X. Zhao}
\affiliation{Beijing National Laboratory for Condensed Matter Physics, Institute of Physics, Chinese Academy of Sciences, Beijing 100190, China.}
\affiliation{School of Physical Sciences, University of Chinese Academy of Sciences, Beijing 100049, China.}
\affiliation{Songshan Lake Materials Laboratory, Dongguan, Guangdong 523808, China}

\author{R. Hackl}\email[Corresponding author: ]{Hackl@wmi.badw.de}
\affiliation{Walther Meissner Institut, Bayerische Akademie der Wissenschaften,
85748 Garching, Germany}
\affiliation{Fakult\"at f\"ur Physik E23, Technische Universit\"at M\"unchen,
85748 Garching, Germany}

\date{\today}
\begin{abstract}
We studied the electronic Raman spectra of \FeSelllll as a function of light polarization and temperature. In the \Blg spectra alone we observe the redistribution of spectral weight expected for a superconductor and two well-resolved peaks below \Tc. The nearly resolution-limited peak at 110\,cm$^{-1}$ (13.6\,meV) is identified as a collective mode. The peak at 190\,cm$^{-1}$ (23.6\,meV) is presumably another collective mode since the line is symmetric and its energy is significantly below the gap energy observed by single-particle spectroscopies. Given the experimental band structure of \FeSelllll, the most plausible explanations include conventional spin-fluctuation pairing between the electron bands and the incipient hole band and pairing between the hybridized electron bands. The absence of gap features in \Alg and \BZg symmetry favors the second case. Thus, in spite of various differences between the pnictides and chalcogenides, this Letter demonstrates the proximity of pairing states and the importance of band structure effects in the Fe-based compounds.

\end{abstract}


\maketitle


The mechanism of Cooper pairing in Fe-based superconductors (FeSCs) or the copper-oxygen compounds is among the most vexing problems in condensed matter physics. The at least partial understanding of these unconventional superconductors would pave the way toward new materials. In either case superconductivity occurs close to magnetic order \cite{Paglione:2010}. Consequently, spin fluctuations are among the candidates for supporting electron pairing \cite{Mazin:2008,Scalapino:2012}. Alternatively, charge \cite{Onari:2009} or orbital fluctuations \cite{Kontani:2010} between the Fe $3d$ orbitals, spin-orbit coupling \cite{Borisenko:2016} and/or nematic fluctuations \cite{Lederer:2015} may support Cooper pairing. In all cases the Fermi surface topology strongly influences the pairing tendencies and qualitative differences between the pnictides and chalcogenides may be expected and were scrutinized in doped BaFe$_2$As$_2$ (122) and FeSe-based (11) compounds.

\begin{figure}[ht!]
  \centering
  \includegraphics[width=8.5cm]{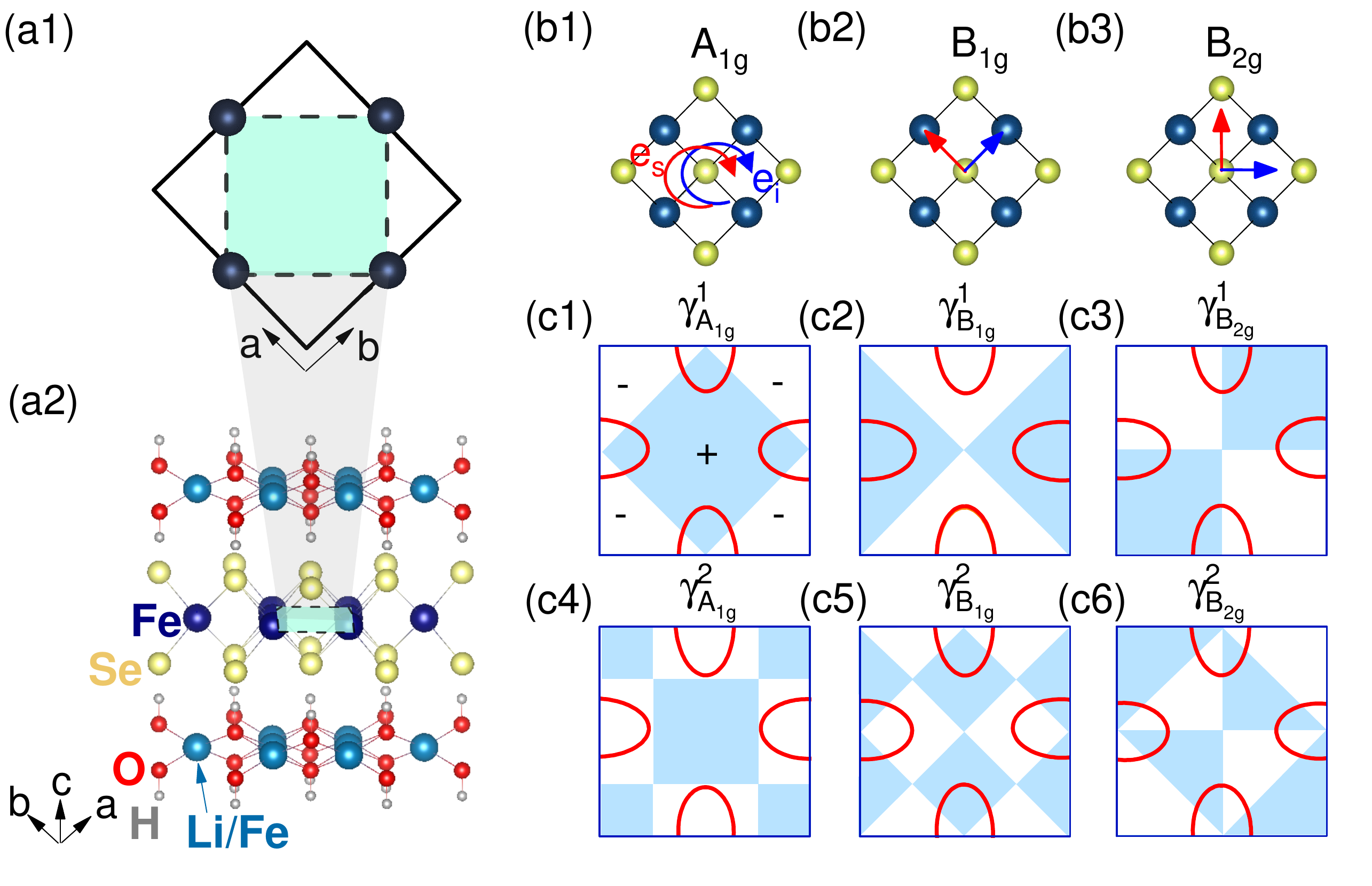}
  \caption{(a1)~1\,Fe (dashes) and 2\,Fe (solid line) unit cells. (a2)~The crystal structure of \FeSelllll. (b1)-(b3) ~ Polarization configurations of \Alg, \Blg and \BZg symmetries. The polarization configurations are indicated on the FeSe-layer. (c1)-(c3) First- and (c4)-(c6) second-order Raman vertices in the first Brillouin Zone (BZ, 1\,Fe unit cell) of  \Alg, \Blg and \BZg symmetry, respectively, for the $D_{4h}$ point group. The unfolded electron pockets are shown as red half ellipses.}
	\label{fig:structure}
\end{figure}

Intercalated FeSe superconductors show \Tc values higher than 40\,K (refs. \cite{Lu:2015,Burrard-Lucas:2013}) but a Fermi surface topology different from the  pnictides. In \FeSelllll, as shown in Fig.~\ref{fig:structure}~(a2), the holelike Fermi surface encircling the $\Gamma$ point in 122 compounds and bulk FeSe cannot be resolved in angle-resolved photoemission spectroscopy (ARPES) any further \cite{Zhao:2016} while it is still present in density functional theory\cite{Nekrasov:2015} marking one of the similarities between intercalated and monolayer FeSe \cite{Zhao:2016,Shi:2017}. This similarity triggers the question as to the pairing interactions and the directly related gap structure. Neither the recent ARPES nor the tunneling experiments yielded clear answers here but show only that there are essentially two rather different gap energies \cite{Du:2016,Du:2018,Chen:2019} on the presumably hybridized concentric electronlike Fermi surfaces \cite{Zhao:2016}.

Here electronic Raman scattering can contribute useful information \cite{Devereaux:2007, Scalapino:2009, Maiti:2016, Kretzschmar:2013, Bohm:2014}. In addition to the gap formation and the pair-breaking peaks at approximately twice the gap energy \cite{Abrikosov:1961, Klein:1984}, collective excitations appear in the Raman response which are related to details of the pairing potential $V_{{\bf k}{\bf k}^\prime}$. Collective excitations in superconductors were first discussed by  Bardasis and Schrieffer (BS) \cite{Bardasis:1961} and by Leggett \cite{Leggett:1966}. The BS mode stems from a subleading pairing interaction that is orthogonal to the ground state. The Leggett mode is best thought of as interband Josephson-like number-phase fluctuation, the absolute energy of which corresponds to the relative coupling strength between the bands in comparison to the intraband coupling \cite{Klein:2010, Blumberg:2007}. Thus, the careful study of putative collective modes offers an opportunity to clarify the competing superconducting instabilities and the related pairing glue.

In this Letter, we present polarization-dependent Raman spectra for temperatures between 7.2\,K and 48\,K in high-quality single-crystalline \FeSelllll thin films with $T_{c} = 42$\,K. At 7.2\,K we observe two well-defined features at 110 and 190\,cm$^{-1}$ in \Blg but not in \Alg and \BZg symmetry. We conclude that at least the resolution-limited line at 110\,cm$^{-1}$ is a collective mode being either related to a subleading pairing interaction or a number-phase oscillation between the electron bands. The superconducting ground state in \FeSelllll may result from either spin fluctuations between the electron bands and the incipient hole band or the interaction between the hybridized electron bands.


\FeSelllll \red{($x\sim0.18$)} thin films were grown epitaxially on $(00l)$-oriented LaAlO$_3$ substrates as reported previously \cite{Huang:2017, Huang:20172}. The thin films have a typical thickness of 100\,nm and were characterized by x-ray diffraction and magnetization measurements showing a high crystallinity and a superconducting transition at $T_c = 42\pm1$\,K (see Supplemental Material A). The Raman experiments were carried out with a standard light scattering equipment \cite{Devereaux:2007}. For excitation we used a solid-state and an Ar$^+$ laser emitting at 577\, and 457\,nm, respectively. All spectra were measured with an absorbed laser power of $P_{\rm abs} = 2$\,mW limiting the heating in the spot to below 1.5\,K/mW (see Supplemental Material B). The polarizations of the incoming and scattered photons will be defined with respect to the 1\,Fe unit cell as shown in Figs.~\ref{fig:structure}(b1)-~\ref{fig:structure}(b3), which are more appropriate for electronic excitations. We show Raman susceptibilities $R\chi^{\prime\prime}(T, \Omega )=S(\Omega, T)\{1+n(T,\Omega)\}^{-1}$ , where $R$ is an experimental constant, $S(\Omega,T)$ is the dynamical structure factor that is proportional to the rate of scattered photons, and $n(T,\Omega)$  is the Bose-Einstein distribution function. The first- and second-order crystal harmonics of each symmetry which Raman vertices are proportional to and the position of the Fermi pockets of \FeSelllll are shown in Figs.~\ref{fig:structure}(c1)-~\ref{fig:structure}(c6) to illustrate the relation between electronic Raman response in different symmetries and the Fermi surface topology.

\begin{figure}[ht!]
  \centering
  \includegraphics[width=9cm]{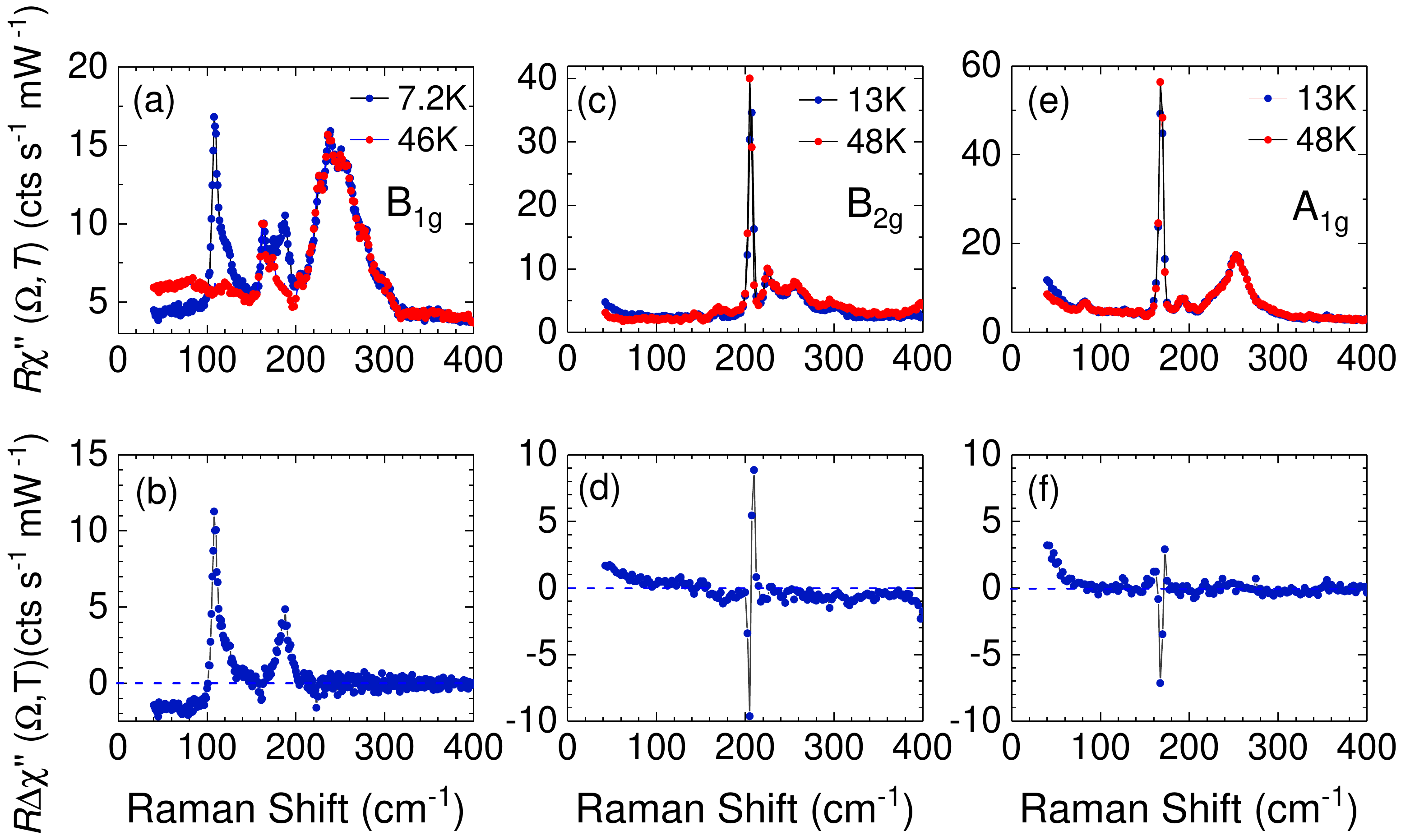}
  \caption{(a),(c),(e)~Raman spectra of \FeSelllll in \Blg, \BZg, and \Alg symmetry. (b),(d),(f)~Difference spectra between superconducting and normal state.}
	\label{fig:results}
\end{figure}

Figure~\ref{fig:results} shows the polarization-dependent Raman response of \FeSelllll above (red) and below (blue) \Tc. The peaks observed at approximately 167\,cm$^{-1}$ in \Alg symmetry [Fig.~\ref{fig:results}~(e)] and 205\,cm$^{-1}$ in \BZg symmetry [Fig.~\ref{fig:results}~(c)] correspond to Se in-phase and Fe out-of-phase vibrations along the $c$ axis, respectively \cite{Zhang:2019}. The peak at 165\,cm$^{-1}$ in the \Blg spectra [Fig.~\ref{fig:results}~(a)] can be identified as a leakage from the \Alg phonon.
The small line width and high intensity of the phonon lines underpin the excellent crystalline quality of the sample. Because of surface contamination, broad peaks at 240\,cm$^{-1}$ [Fig.~\ref{fig:results}~(a)], 224 and 256\,cm$^{-1}$ [Fig.~\ref{fig:results}~(c)], and 252\,cm$^{-1}$ [Fig.~\ref{fig:results}~(e)] in \Blg, \BZg and \Alg symmetry, respectively, appear in the spectra. These peaks probably originate from Fe oxide phonons since they disappear after cleaving and are discussed in more detail in Supplemental Material C.

If the normal state spectra are subtracted from the superconducting spectra all phonons and extra lines disappear since they do not change appreciably upon crossing \Tc as shown in Figs.~\ref{fig:results}~(b), ~\ref{fig:results}(d), and ~\ref{fig:results}(f). However, due to the high spectral resolution even small changes of the phonon lines can be identified.

The small changes in the phonon lines are the only detectable effects of superconductivity in \Alg and \BZg symmetry. The increase toward zero energy is an artifact resulting from insufficient rejection of the laser line. The spectral changes in \Blg symmetry are resolved clearly since the extra peaks at approximately 110\,cm$^{-1}$ and 190\,cm$^{-1}$ have an intensity comparable to that of the phonons. In addition to the peaks, the continuum is suppressed below 90\,cm$^{-1}$ and is nearly energy independent. None of the excitations display appreciable resonance behavior, and the data can be reproduced in different regions of the sample, as shown in Supplemental Material D and E. The suppression and the additional peaks indicate a relation to superconductivity.

The most important observations include the following:\\
(i) There is no intensity redistribution below \Tc in \Alg and \BZg symmetry, as observed earlier\cite{Kretzschmar:2013,Muschler:2009}. The phenomenon can be understood qualitatively in terms of the related polarization-dependent Raman form factors, the Fermi surface topology of \FeSelllll as shown in Figs. ~\ref{fig:structure}(c1) and ~\ref{fig:structure}(c3), and screening effects.\\
(ii) The\Blg spectrum is suppressed below 90\,cm$^{-1}$. This indicates a nearly isotropic superconducting gap as already observed for ${\rm Rb_{0.8}Fe_{1.6}Se_2}$ \cite{Kretzschmar:2013}. The residual intensity of approximately 2-3\,counts(s mW)$^{-1}$ is not entirely clear but may either originate from the substrate, surface layers, or luminescence. In agreement with the scanning tunneling spectroscopy (STS) results \cite{Du:2016}, there is no reason to assume that there are states inside the gap. Similar residual Raman intensities are also observed in single crystals of pnictides and other chalcogenides \cite{Kretzschmar:2013, Bohm:2014, Jost:2018} and can safely be assumed to be extrinsic.\\
(iii) There are two superconductivity-induced features separated by some 80\,cm$^{-1}$. Gap features at a similar separation but slightly higher energies were observed by STS and ARPES as summarized in Table~\ref{tab:energies}. The significant differences in the derived gap energies presumably have a real physical reason such as a substantial energy difference between the gap and collective modes \cite{Bohm:2018}. The line at 190\,cm$^{-1}$ is nearly symmetric and 10-12\wn wide while that at 110\,cm$^{-1}$ is rather sharp around the maximum but asymmetric.\\
(iv) The peak at 110\wn is nearly resolution-limited. For its width, it cannot result from pair breaking alone. Rather the symmetric line at 110\,cm$^{-1}$ having a FWHM of less than 5\,cm$^{-1}$ is a collective mode while the shoulder on the high-energy side originates  from pair breaking. We will explore this possibility later by a phenomenological analysis.

\begin{figure}[ht!]
  \centering
  \includegraphics[width=5cm]{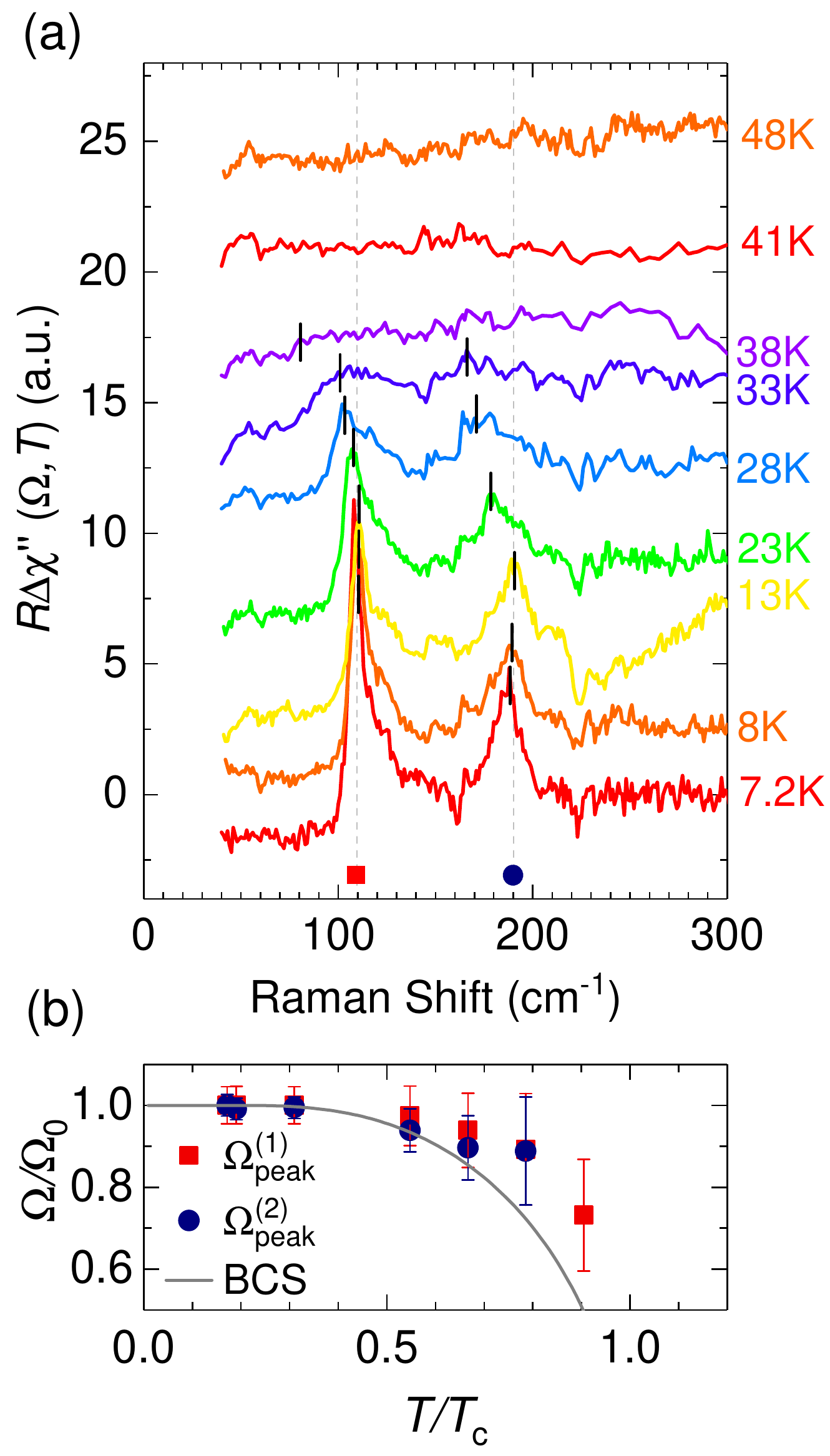}
  \caption{(a)~ Temperature dependence of the difference spectra in \Blg symmetry. For clarity data other than those measured at the lowest temperature are shifted vertically. The peak positions are marked by the black vertical lines. Two dashed lines indicate the peak positions at 7.2\,K. The increase of the 13 K spectrum toward higher energies originates from a local surface contamination. (b)~Normalized temperature dependence of peak positions. They are extracted by a fit with a Lorentzian function. For normalization of the energy the  $\Omega_0$ = 110 and 190\,cm$^{-1}$ for peak 1 and peak 2, respectively.}
	\label{fig:T}
\end{figure}

\begin{table*}
  \caption{Comparison between the peak positions in the \Blg Raman spectra and the gap energies obtained by STS and ARPES.
  }
  \setlength{\tabcolsep}{4mm}{
  \begin{tabular}{ c c c  c c c }
	\specialrule{0em}{2pt}{2pt}
	  \hline
	\specialrule{0em}{2pt}{2pt}
	    &Raman~(cm$^{-1}$) & Raman~(meV) & & STS~(meV) \cite{Du:2016,Chen:2019} & ARPES~(meV) \cite{Zhao:2016} \\
	\specialrule{0em}{2pt}{2pt}
	  \hline
	  \hline
	\specialrule{0em}{1.5pt}{1.5pt}
	  $\Omega_{\rm peak}^{(1)}$ & 110$\pm$0.5 & 13.75$\pm$0.06 & $2\Delta_{1}$& 17.2$\pm$2.0 & -- \\
	\specialrule{0em}{1.5pt}{1.5pt}
	   $\Omega_{\rm peak}^{(2)}$ & 190$\pm$0.5 & 23.75$\pm$0.06 & $2\Delta_{2}$& 28.4$\pm$4.0 & 26.0$\pm$4.0  \\
	\specialrule{0em}{1.5pt}{1.5pt}
	\specialrule{0em}{1.5pt}{1.5pt}
	   $\Omega_{\rm peak}^{(1)}/k_{B}T_{c}$& \multicolumn{2}{c}{3.80$\pm$0.02} & $2\Delta_{1}/k_{B}T_{c}$& 4.75$\pm$0.55 & --  \\
	\specialrule{0em}{1.5pt}{1.5pt}
	\specialrule{0em}{1.5pt}{1.5pt}
	   $\Omega_{\rm peak}^{(2)}/k_{B}T_{c}$& \multicolumn{2}{c}{6.56$\pm$0.02} & $2\Delta_{2}/k_{B}T_{c}$& 7.84$\pm$1.10 & 7.18$\pm$1.10  \\
	\specialrule{0em}{1.5pt}{1.5pt}
	  \hline
  \end{tabular}}
	\label{tab:energies}
\end{table*}

In Fig.~\ref{fig:T}, we show the variation with temperature of the \Blg difference spectra. With increasing temperature, the two peaks shift to lower energy and cannot be resolved any further above 38\,K. Both peaks depend more weakly on temperature than expected from the BCS theory. As opposed to the results in \BKFA \cite{Bohm:2014} this temperature dependence does not allow us to clearly identify the origin of the superconducting structures. Whereas the pair-breaking peaks do not necessarily follow the BCS prediction, except in weakly interacting systems \cite{Devereaux:1995}, at least BS collective modes are expected to follow the related single-particle gap $\Delta(T)$ \cite{Monien:1990}. The temperature dependence of Leggett modes has not been analyzed yet, but is presumably more complicated since the coupling of at least two gaps has to be considered \cite{Suhl:1959,Khodas:2012}. Thus, the variation with temperature is not an identification criterion.

\begin{figure}[ht!]
  \centering
  \includegraphics[width=9cm]{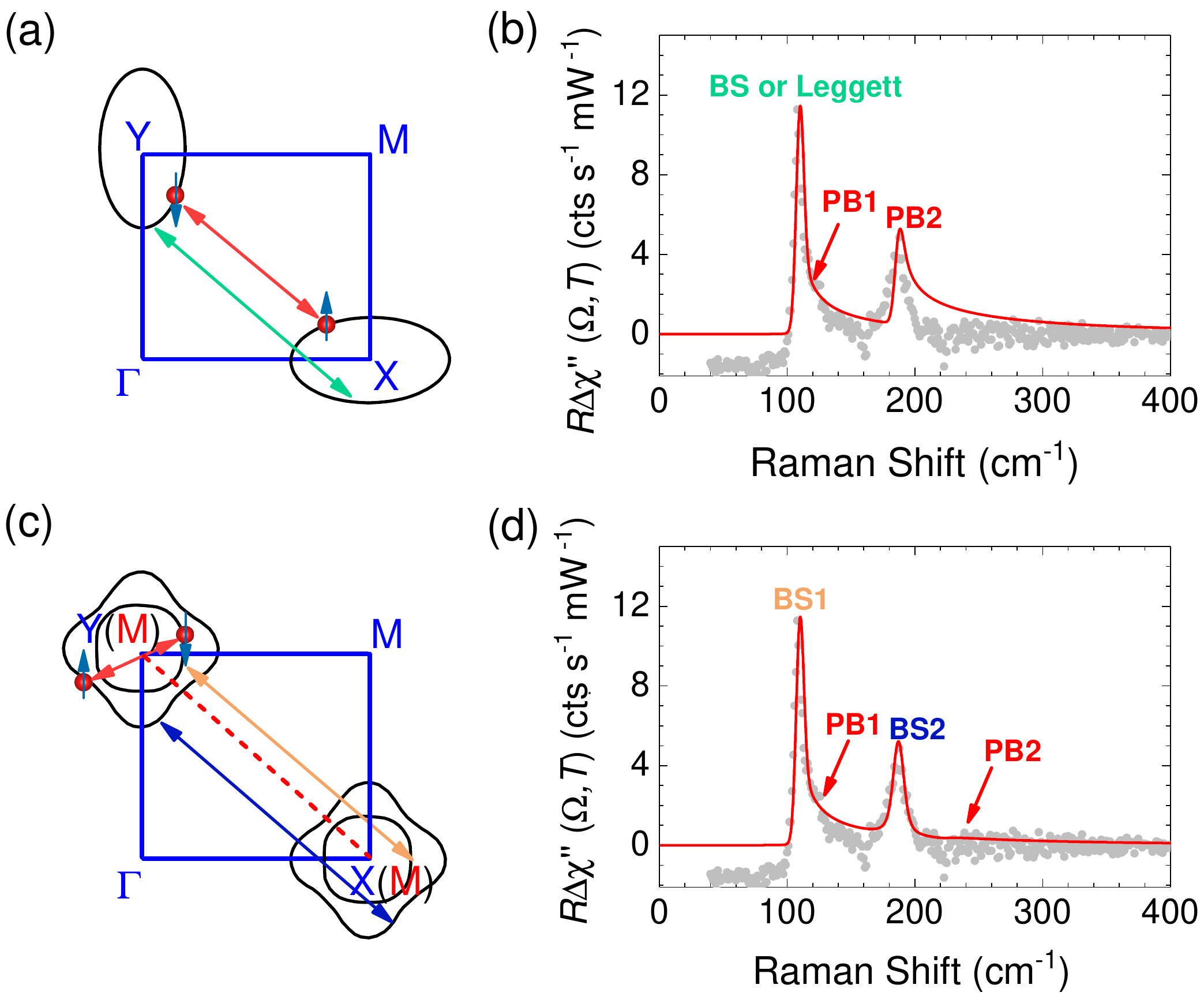}
  \caption{Interactions and related phenomenological spectra. (a),(c) Illustration of the superconducting ground state and possible subdominant interactions in the first BZ. The blue solid lines indicate the first quadrant of the 1 Fe BZ boundary in (a) and (c). The 2 Fe BZ boundary in (c) is illustrated by the red dashed line. The Fermi pockets are shown by the black solid curves. The red double arrows indicate the pairing interaction in (a) and (c). The green double arrow in (a) and the orange and dark blue ones in (c)  indicate the possible subdominant pairing interactions. (b), (d) The fitting results correspond to the situations in (a) and (c). The red curves are the fitting results. The gray solid dots are the experimental data. The BS modes or Leggett mode and pair-breaking (PB) peaks are indicated in (b) and (d).
  }
	\label{fig:pheno}
\end{figure}

What are the possible explanations and the implications thereof for superconductivity in \FeSelllll? \red{Given the STS and ARPES results a scenario with two isotropic $s$-wave gaps appears to be natural and compatible with the Raman spectra. However, the gap energies observed by Raman scattering are significantly too small (Table~\ref{tab:energies}), and the phenomenology returns poor agreement with the data (see Supplemental Material G) }. \red{Rather,} the shape of the line at  110\,cm$^{-1}$ \red{is strongly indicative of} a collective mode. An undamped quadrupolar excitation inside the gap \cite{Thorsmolle:2016} \red{or a nematic resonance \cite{Gallais:2016,Kang:2016} are} not very likely since the related fluctuations above \Tc could not be observed \red{(see Supplemental Material F)}. The distinction between a BS and a Leggett mode is less obvious. Since the mode is apparently below the edge of the smaller gap, the damping is small in either case.

A BS mode \red{indicates} orthogonal pairing channels \cite{Bardasis:1961}. Given the possible interactions between the two electron bands at the $X$ and $Y$ points, as shown in Fig.~\ref{fig:pheno}\,(a), only a $d_{x^2-y^2}$ gap of lowest (first) order is possible if spin fluctuations are relevant. In this case the gaps on the two bands have opposite sign and, since there are no nodes on the Fermi surface, the gap is clean in agreement with the STS and Raman results. Then the BS mode would result from a $d_{x^2-y^2}$ interaction of second or higher order [for second-order \Blg symmetry, cf. Fig.~\ref{fig:structure}(c5)], having a smaller coupling strength than the ground state. The description of the asymmetric maximum at 110\,cm$^{-1}$ is reasonable, as shown in Fig.~\ref{fig:pheno}~(b), and yields a coupling strength of approximately 0.16 for the subleading channel (see Supplemental Material G).

This interpretation favors pair breaking as a possible explanation for the high-energy peak at 190\,cm$^{-1}$. However, the symmetric shape is  untypical for a pair-breaking feature and may only be explained by a broader gap distribution as suggested by the STS data for instance (see Table \ref{tab:energies}). Yet, the energy of 190\,cm$^{-1}$ (23.6\,meV) is significantly below the single-particle gap. More importantly, it is difficult to explain why there should be two rather distinct gaps for this scenario of equivalent bands. It is, in fact, more likely that there is a $(\pi,\pi)$ reconstruction of the Fermi surface and a hybridization between the electron bands, as suggested by Khodas \etal \cite{Khodas:2012,Khodas:2014}. Then one expects two concentric Fermi surfaces with distinctly different gaps as seen here in the Raman data and also in the STS data\cite{Du:2016,Chen:2019}. ARPES \cite{Zhao:2016} and STS \cite{Du:2016} tell us that the outer Fermi surface has the larger gap. Superconductivity would then arise from the comparably strong interaction between these hybridized bands, inducing a repulsion of the gap energies \cite{Du:2018}  and either a collective Leggett mode in the \Alg channel \cite{Cea:2016} or a double-peak structure well below the gap for certain parameter ranges in the case of a sign change of the gap between the bands ($s_\pm$ gap) \cite{Khodas:2014}.

Since only the \Blg channel displays a distinct redistribution of spectral weight below \Tc, scenarios which include other channels are less likely to explain the results. Thus, in addition to the strongly coupled ground state resulting from the interaction between the concentric bands there must be a weaker $(\pi,\pi)$ interaction leading to collective modes in \Blg symmetry, as shown in Fig.~\ref{fig:pheno}(c). In this scenario, the sharp mode is either a Leggett mode from a weak $(\pi,\pi)$ coupling between $X$ and $Y$ on top of the strongly coupled ground state resulting from the strong coupling of the concentric bands or a BS mode having a similar origin. Huang \etal \cite{Huang:2018} indeed argue that the distinction between Leggett and BS modes become\red{s} obsolete here.

The only remaining issue concerns the positions of the Raman peaks that appear at smaller energies than in the single-particle spectroscopies. Whereas the energy of the Raman maximum at 110\,cm$^{-1}$ is naturally explained in terms of a BS mode appearing below the related single-particle gap at 136\wn (17\,meV) and manifesting itself as a shoulder in the Raman spectrum\red{,} the position of the mode at 190\,cm$^{-1}$ is less obvious since there is no additional pair-breaking feature in the spectra. In principle it could be another BS mode pulled down by 10\% from the gap edge at approximately 220\wn (27\,meV) and suppressing the pair breaking almost entirely. Also in this case the ground state would be induced by a strong hybridization of the two electron bands. The resulting description of the experimental data is in fact much better in this case [see Fig.~\ref{fig:pheno}~(d)].

Finally, since the hole band is rather close to the Fermi surface, spin-fluctuation pairing between the incipient  hole band and the electron bands can still be rather strong \cite{Linscheid:2016,Mishra:2016}. Then the ground state is $s_\pm$ with all the electron bands having the same sign, and the \Blg modes are collective $d$-wave modes from the hybridized electron bands. Whether or not the magnitudes of the gaps observed on the electron bands and the related subleading pairing strengths are compatible with these considerations needs to be worked out theoretically. The resulting spectra could be very similar to those in the previous case. However, due to the pairing-induced renormalization of the central hole band below \Tc  one would not expect the \Alg spectra to be entirely insensitive to superconductivity. Thus the scenario of an incipient band is less supported by the present experiment.


In conclusion, we studied the polarization- and temperature-dependent Raman spectra in \FeSelllll. Superconductivity affects only the \Blg spectra. One of the observed modes is resolution limited arguing strongly for its collective character. For the surprisingly successful description of the data in terms of BS modes there are essentially two scenarios: (i) dominant pairing between the hybridized electron bands \cite{Khodas:2012} and subleading ($\pi$, $\pi$) interactions between the electron bands. Then Leggett and BS modes cannot be distinguished \cite{Huang:2018}. (ii) If the ground state originates from the interaction between the incipient hole band and the electron bands \red{a} similar collective mode may be expected\red{,} but completely inert \Alg spectra are unlikely in this case making scenario (i) more likely. Yet, the distinction between the two scenarios requires quantitative theoretical studies.

\begin{acknowledgments} We thank L. Benfatto, P. Hirschfeld, T. Maier and L. Zhao for fruitful discussions. This work is supported by the Deutsche Forschungsgemeinschaft (DFG) through the coordinated programme TRR80 (Projekt-ID 107745057) and project HA2071/12-1. G. H. would like to thank the Alexander von Humboldt Foundation for support from a research fellowship.  The work at China was supported by National Natural Science Foundation of China (No. 11834016), and the National Key Research and Development Program of China (Grant No. 2017YFA0303003) and Key Research Program of Frontier Sciences of the Chinese Academy of Sciences (Grant No. QYZDY-SSW-SLH001).
).

\end{acknowledgments}

\section*{Supplementary information}
\noindent \textbf{A. Sample characterization}
~\\

\begin{figure}[ht!]
\centering
	\includegraphics[width=6.5cm]{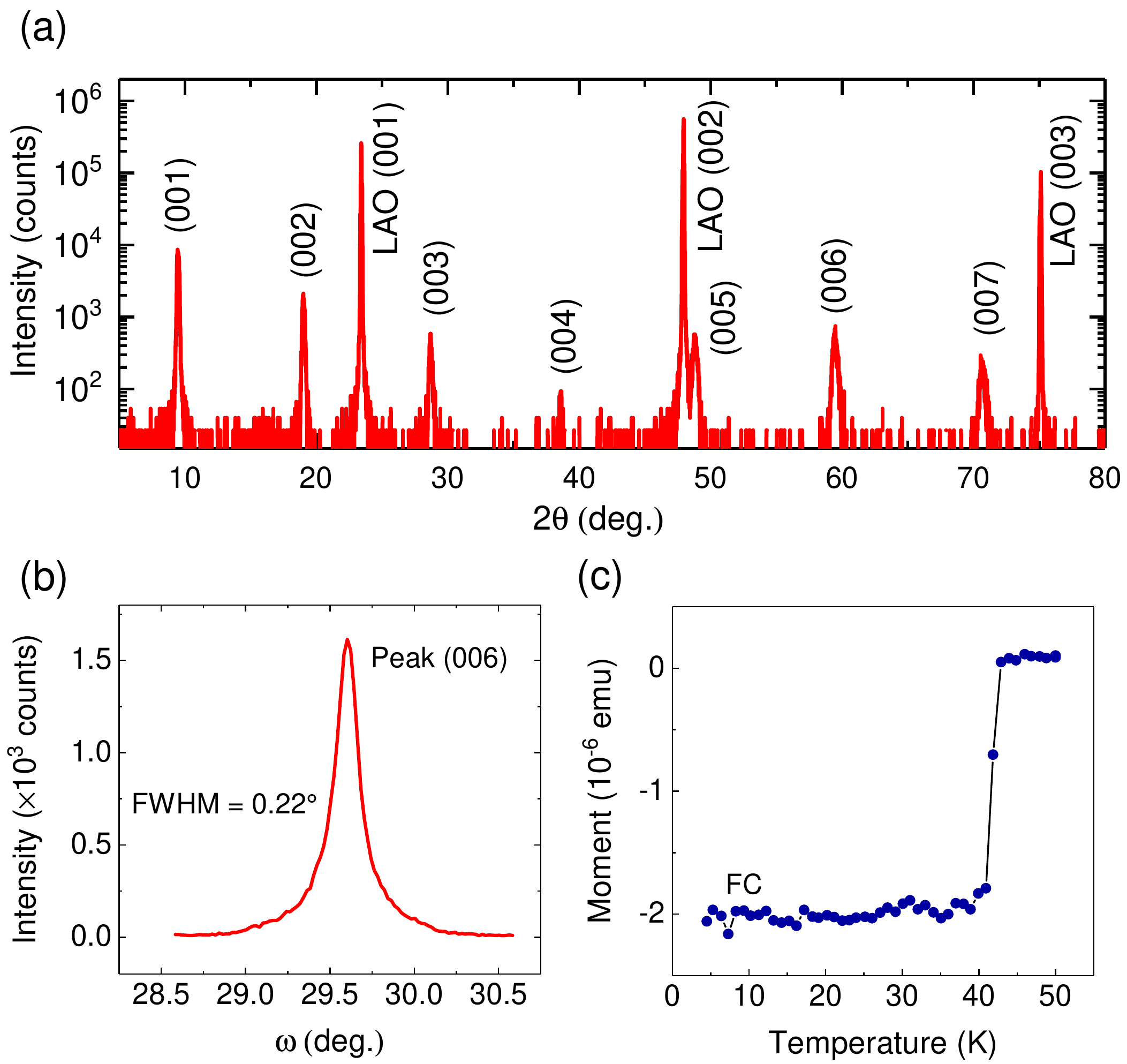}
	\caption{ \red{C}haracterizations of the (Li$_{1-x}$Fe$_{x}$)OHFeSe film on LaAlO$_3$ substrate. (a)~$\theta-2\theta$ scan. \red{The scan} shows only (00$l$) peaks. \red{The LAO peaks are indicated.} (b)~\red{R}ocking curve of (006) reflection. \red{The} FWHM \red{is as small as} 0.22$^\circ$. (c)~Temperature dependence of field-cooling (FC) magnetization \red{with the magnetic field of 1 Oe}. \red{T}he \red{diamagnetic transition} occur\red{s} at $\sim$42\red{$\pm1$}\,K.}
	\label{figS1}
\end{figure}

X-ray diffraction (XRD) measurements were carried out on a 9\,kW Rigaku SmartLab x-ray diffractometer. Magnetization measurements were done using a Quantum Design MPMS-XL1 system with a measuring field of 1 Oe to characterize the superconducting state. Figure~\ref{figS1}~(a) and (b) show the XRD characterizations of the (Li$_{1-x}$Fe$_{x}$)OHFeSe films. The XRD pattern exhibits a single preferred orientation of (00$l$) for (Li$_{1-x}$Fe$_{x}$)OHFeSe, and the peaks of LaAlO$_3$ substrates are marked with LAO. The corresponding full width at half maximum (FWHM) of the rocking curve for the (006) reflection is 0.22$^\circ$, indicative of the high crystalline quality. Figure~\ref{figS1}~(c) presents the typical temperature dependent magnetization of \red{a small piece of a}  (Li$_{1-x}$Fe$_{x}$)OHFeSe film.  \red{Sharp transition was reproduced for several small pieces (approximately 2$\times$2\,mm$^2$) of thin films of the same batch and indicates that the sample is homogeneous.} The $T_c$, obtained from the onset temperature of diamagnetism, of optimal (Li$_{1-x}$Fe$_{x}$)OHFeSe films is 42\,K. \red{The transition width (10 to 90\,\% of the full signal) is less than 2\,K.} \red{The film thickness is on the order of 100\,nm as reported in Ref. \cite{Huang:2017}}.

\noindent \textbf{B. Determination of the spot temperature}
~\\

\begin{figure}[ht!]
	\centering
	\includegraphics[width=7.5cm]{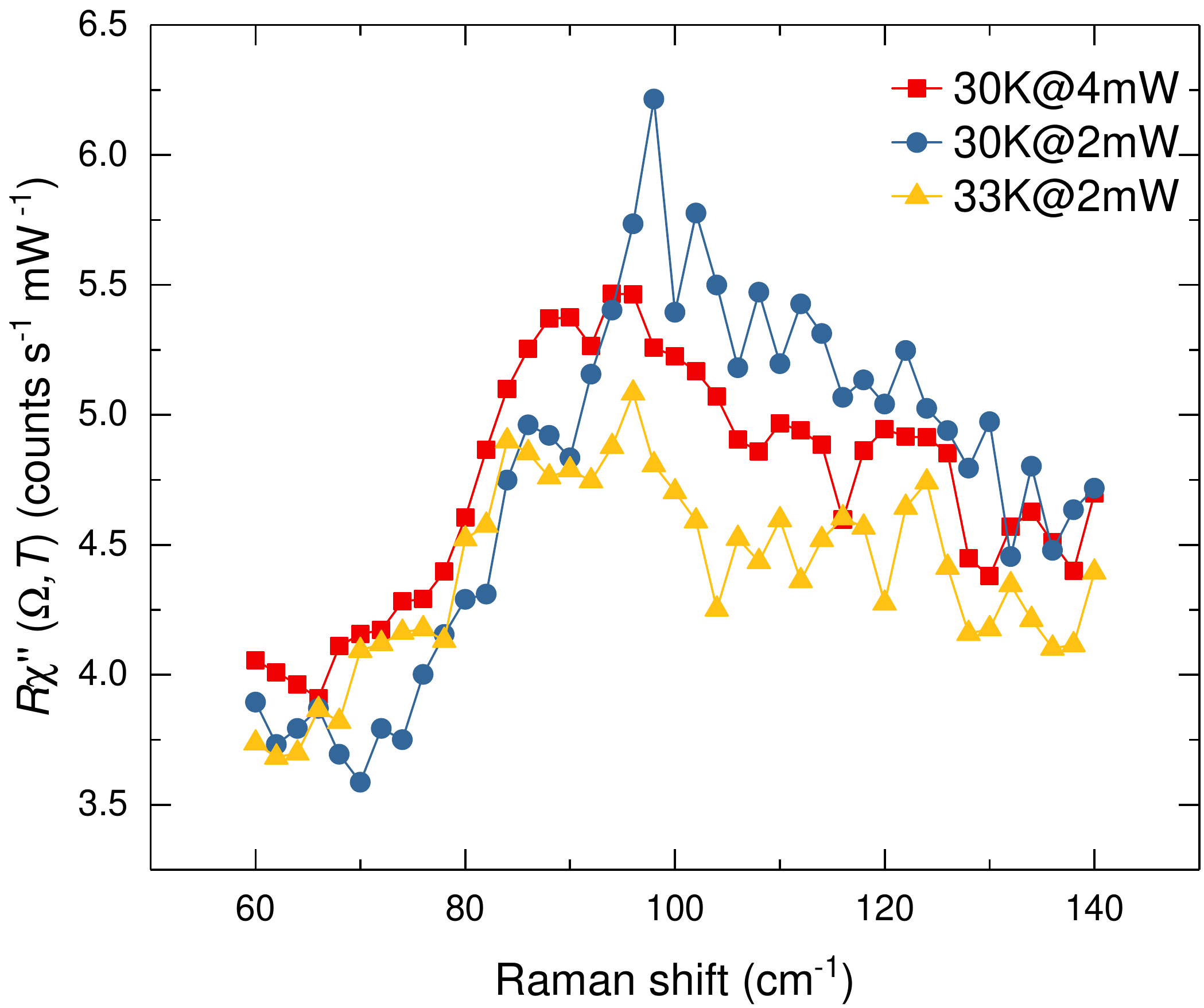}
	\caption{Determination of the spot temperature. The spectra were obtained in \Blg symmetry.}
	\label{figS2}
\end{figure}

The real temperature in the illuminated spot is higher than the holder temperature due to the laser heating effect. In order to figure out the heating, we measure the spectra in \Blg symmetry at 30\,K and 33\,K with different laser power. As seen in Fig.~\ref{figS2}, the peak positions \red{measured} at 30\,K \red{with an absorbed laser power of }4\,mW and \red{at} 33\,K \red{measured with }2\,mW are almost the same. Therefore, we can estimate that the laser induced heating is approximately 1.5\,K/mW.
~\\

\noindent \textbf{C. Surface quality}
~\\

\begin{figure}[ht!]
	\centering
	\includegraphics[width=7.5cm]{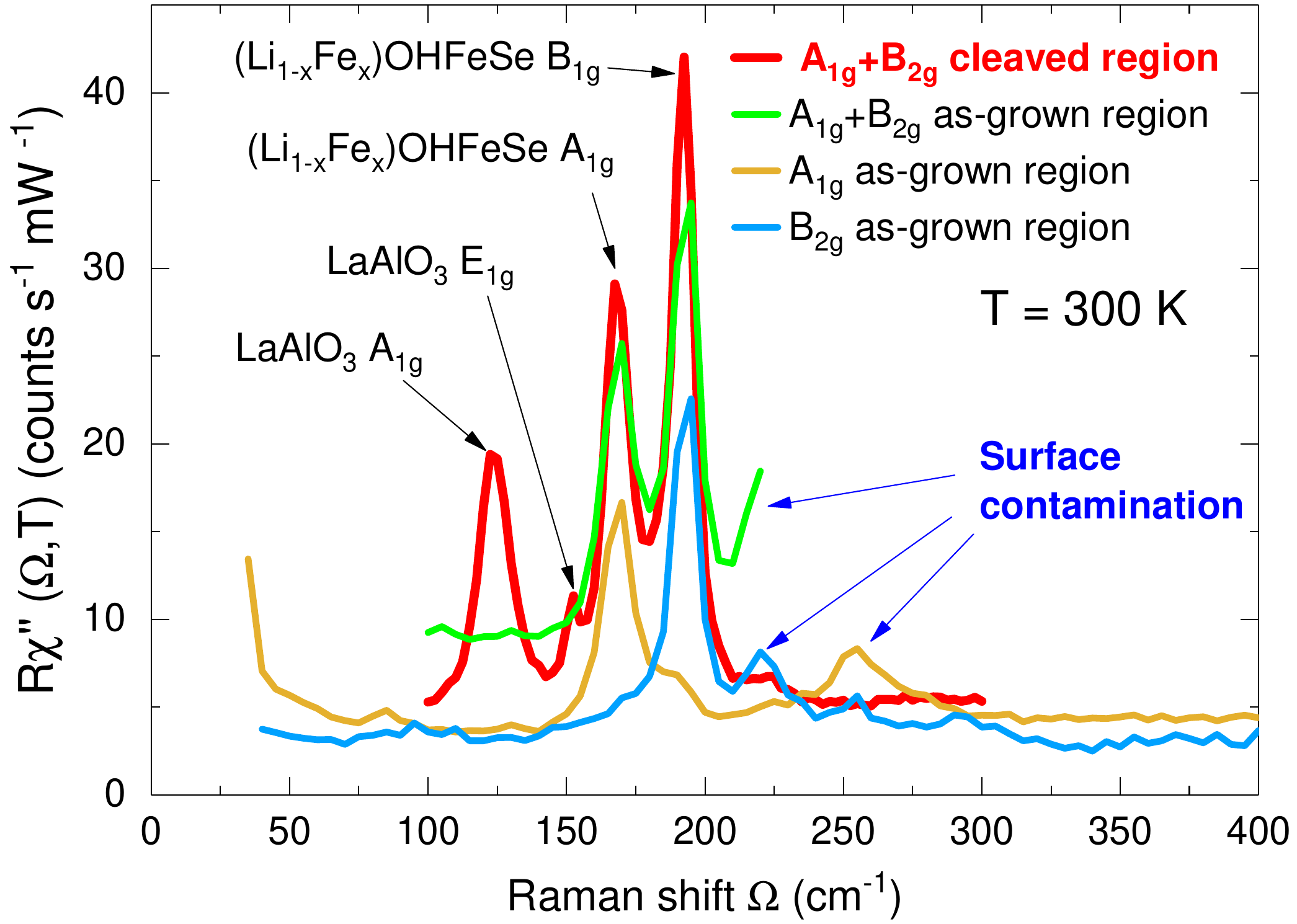}
	\caption{ Raman spectra in cleaved and as-grown regions of the (Li$_{1-x}$Fe$_{x}$)OHFeSe sample. The \Alg and \Eg phonons of the substrate (LaAlO$_3$) can be observed in the cleaved region. The \Alg and \Blg phonons of (Li$_{1-x}$Fe$_{x}$)OHFeSe sample are identified in both cleaved and as-grown region. The blue arrows indicate the additional peaks induced by surface contamination.}
	\label{figS3}
\end{figure}

To clarify the origin of the additional peaks appearing in the range of 220-260 cm$^{-1}$ in \Alg, \Blg and \BZg symmetry, we compared the spectra in cleaved and as-grown regions as shown in Fig.~\ref{figS3}. It can be clearly seen that the additional peaks disappear in the cleaved region in both \Alg and \BZg symmetry. It means that they stem from surface contamination or degradation, and originate probably from the \Alg, \Eg (1), \Eg (2) and \Eg (3) phonons of Fe oxides \cite{Shim:2002}. We can rule out a luminescence effect since these peaks stay pinned and only change intensity for different laser lines (see SI.D) as shown in Fig.~\ref{figS5}. Because the film is too thin, the \Alg and \Eg phonons of the substrate (LaAlO$_3$) \cite{Sathe:2007} become stronger after cleavage.
~\\

\noindent \textbf{D. Resonances}
~\\

\begin{figure}[ht!]
	\centering
	\includegraphics[width=7.5cm]{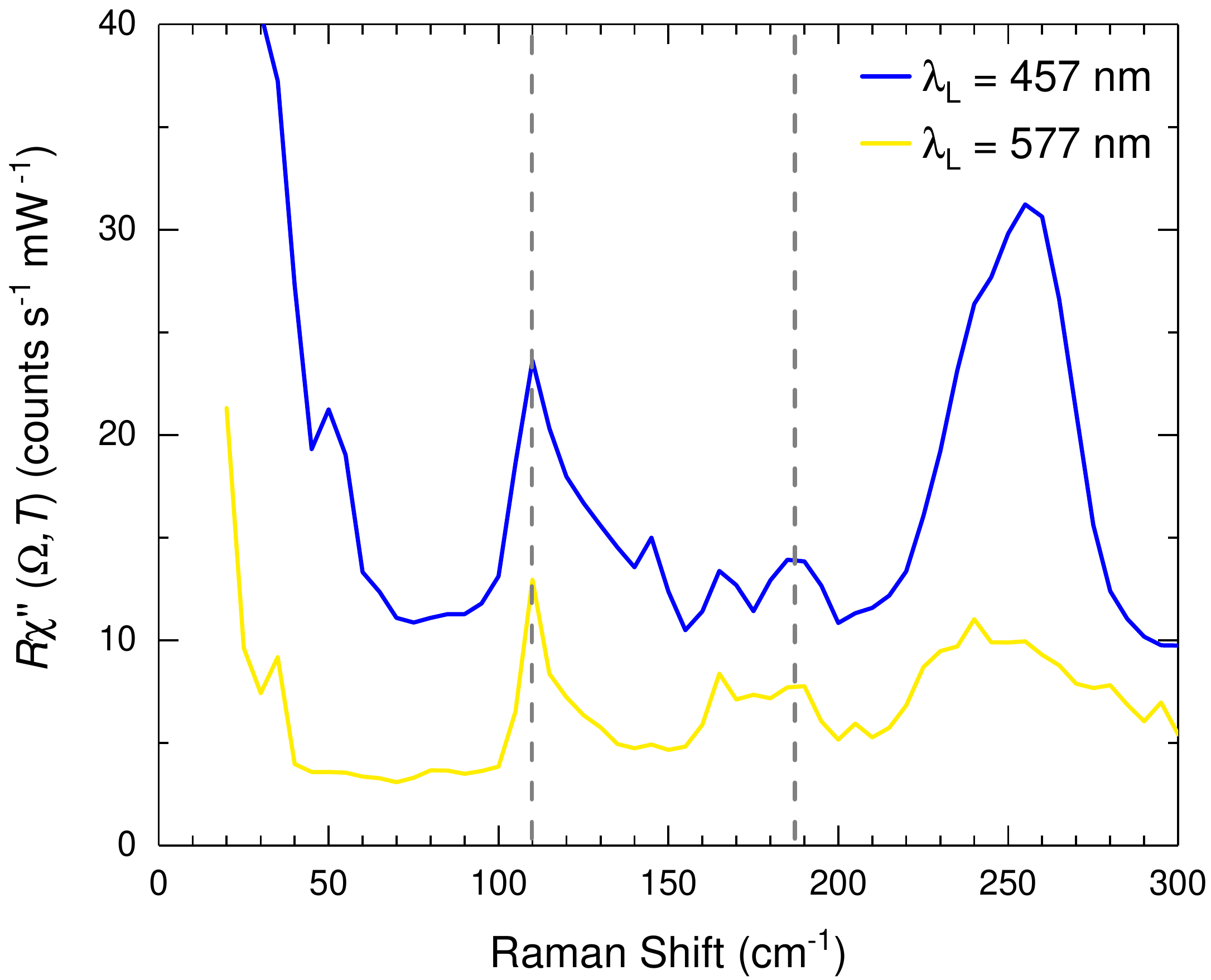}
	\caption{ Comparison between blue laser excitation and yellow laser excitation in \Blg symmetry at 7.2\,K and 2\,mW. }
	\label{figS4}
\end{figure}

\begin{figure}[ht!]
	\centering
	\includegraphics[width=7.5cm]{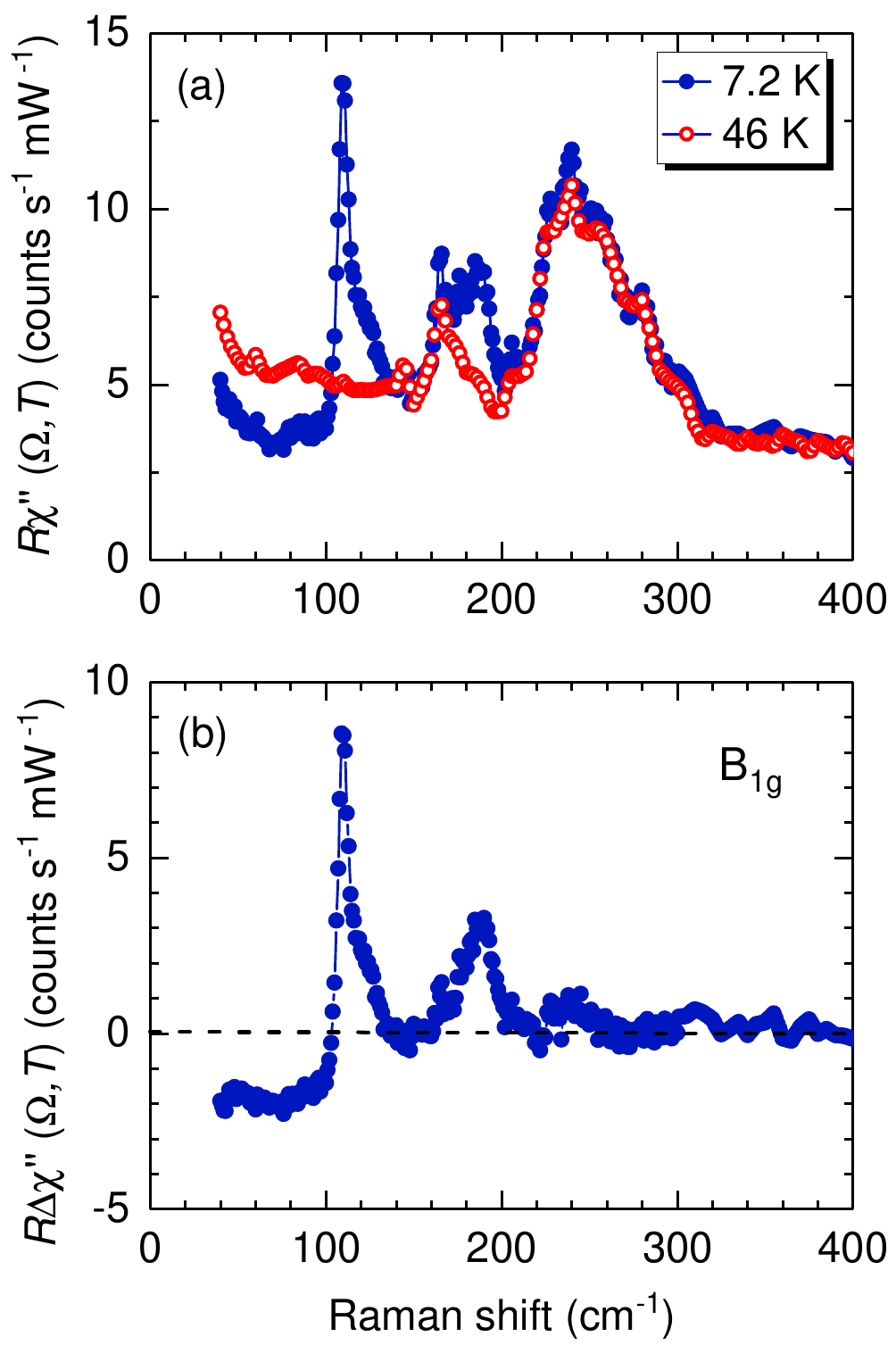}
	\caption{ Superconducting and normal state Raman spectra of (Li$_{1-x}$Fe$_{x}$)OHFeSe thin film in \Blg channel taken from the second spot. (a) Raman spectra at 7.2 K and 46 K. (b) The difference spectra between 7.2 K and 46 K, i.e. $R\chi^{''}(\Omega, 7.2 K)-R\chi^{''}(\Omega, 46 K)$.}
	\label{figS5}
\end{figure}

Figure~\ref{figS4} shows the Raman spectra in \Blg symmetry for the excitation lines at 457 and 577\,nm. The peak at 250\,cm$^{-1}$ is enhanced for blue laser excitation. The peak at 50\,cm$^{-1}$ is a plasma line in the blue. The peak at 35\,cm$^{-1}$ with yellow excitation, is of unknown origin. However, neither of them is related to superconductivity since they also appear above the superconducting transition temperature. The superconductivity-induced peaks are located at the same positions for both yellow and blue laser (see dashed lines in Fig.~\ref{figS5}), which means that resonance effects play a minor or no role.
~\\

\begin{figure}[ht!]
	\centering
	\includegraphics[width=7cm]{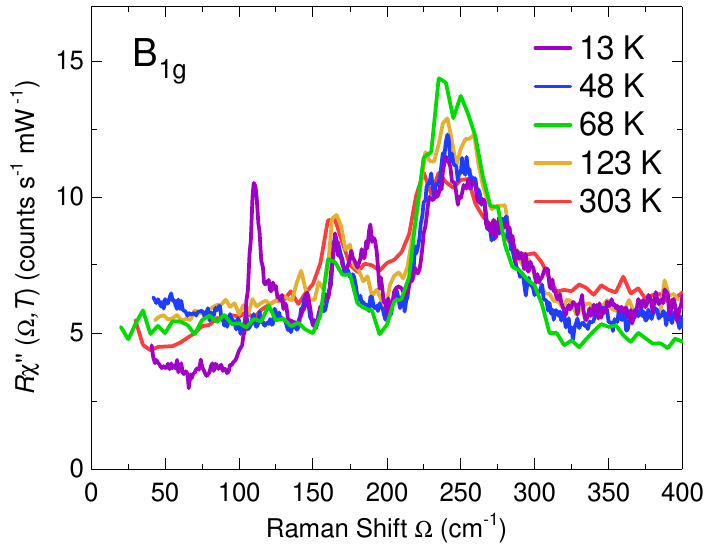}
  \caption{\rev{Raman spectra in \Blg symmetry at various temperatures. Partially due to the limited rejection of stray light fluctuations cannot be detected in (Li$_{1-x}$Fe$_{x}$)OHFeSe}.}
	\label{figS6}
\end{figure}
\noindent \textbf{E. Data reproducibility}
~\\

Figure~\ref{figS5} shows the Raman spectra in \Blg symmetry taken from the second spot on the same sample. The two superconductivity-induced peaks can be observed at 110 and 190\,cm$^{-1}$, which are consistent with the results of the first spot (Fig. 2 in the main text). The peaks at 165\,cm$^{-1}$ and 240\,cm$^{-1}$ are ascribed to \red{a leakage of} the \Alg phonon and phonons of Fe oxides respectively, as described in the main text and SI. C. All these features indicate that our data \red{are} well reproduc\red{ible}.
~\\


\noindent \textbf{\rev{F. Fluctuations}}
~\\

\rev{Gallais \etal proposed that nematic fluctuations may enhance the pair-breaking peak resonantly \cite{Gallais:2016}. In fact, if there are fluctuations in the system, an extra contribution may be superimposed on the electronic continuum in the Raman spectra \cite{Kretzschmar:20162}. This contribution cannot be observed in the \Blg Raman spectra of (Li$_{1-x}$Fe$_{x}$)OHFeSe above \Tc as shown in Fig.~\ref{figS6}. One could argue that the low-energy part is not as well resolved in (Li$_{1-x}$Fe$_{x}$)OHFeSe as in \BFCA as a consequence of the poor stray-light rejection. Yet, it is fair to say that there is no clear evidence for temperature dependence comparable to that in \BFCA (see, e.g. our recent paper in Phil. Mag. \cite{Lederer:2020}). In addition, the spectra in (Li$_{1-x}$Fe$_{x}$)OHFeSe and optimally doped \BFCA look very different, and, although more work on (Li$_{1-x}$Fe$_{x}$)OHFeSe samples having better surfaces is necessary, we tend to conclude that fluctuations cannot be observed in (Li$_{1-x}$Fe$_{x}$)OHFeSe above \Tc. We note that the line shape of a nematic resonance cannot \textit{a priory} be distinguished from a BS mode on the basis of just one doping level.}
~\\
\begin{table*}[ht!]
          \caption{Fitting parameters of difference spectra in \Blg symmetry.}
	\setlength{\tabcolsep}{2.5mm}{
	\begin{tabular}{c c c c c c c c c c}
	\specialrule{0em}{2pt}{2pt}
	  \hline
	\specialrule{0em}{2pt}{2pt}
	   Parameters &2$\Delta_0^1$(cm$^{-1}$) &2$\Delta_0^2$(cm$^{-1}$) & $ \lambda_d^1$ & $\lambda_d^2$ & $\lambda_s$ & $\delta$ & $b_1$ & $b_2$ & $\sigma$(cm$^{-1}$)\\
	\specialrule{0em}{2pt}{2pt}
	  \hline
	  \hline
	\specialrule{0em}{1.5pt}{1.5pt}
	BS or Leggett mode + PB peak &  112 & 186 & 0.16 & 0 & 1 & 0.1& 0.53 & 0.30 & 3\\
	\specialrule{0em}{1.5pt}{1.5pt}
	\specialrule{0em}{1.5pt}{1.5pt}
	Two BS  modes & 112 & 224 & 0.16 & 0.40 & 1 & 0.1 & 0.53 & 0.12 & 3 \\
	\specialrule{0em}{1.5pt}{1.5pt}
	\specialrule{0em}{1.5pt}{1.5pt}
	\rev{Two PB peaks} & \rev{108} & \rev{186} & \rev{0} & \rev{0} & \rev{1} & \rev{0.1} & \rev{0.82} & \rev{0.20} & \rev{3} \\
	\specialrule{0em}{1.5pt}{1.5pt}
	  \hline
	\end{tabular}}
	\label{tab1}
\end{table*}


\begin{figure}[ht!]
	\centering
	\includegraphics[width=6cm]{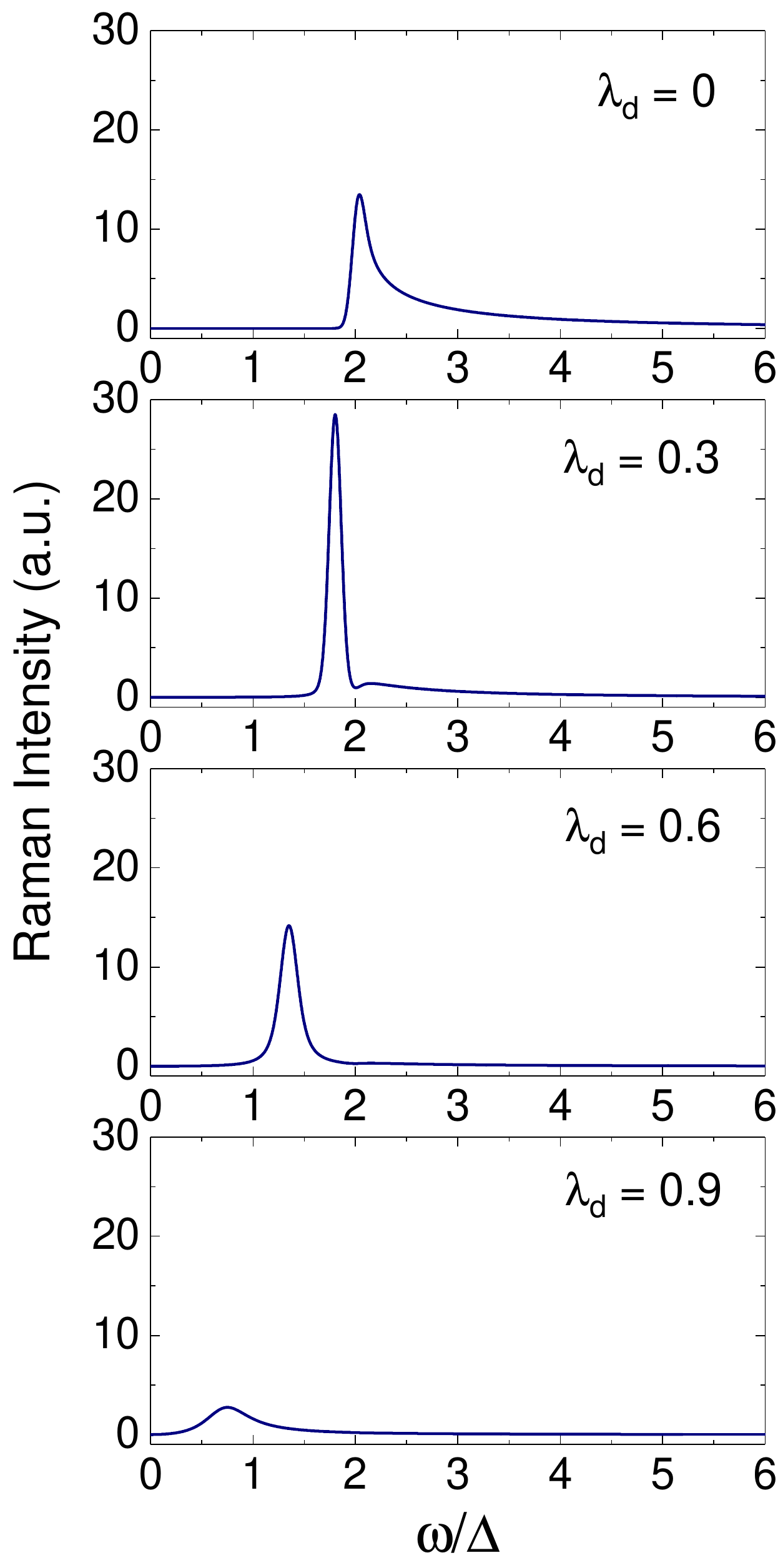}
	\caption{ Plots of the Raman response for different coupling strength $\lambda_d$. Here the parameters are set as follows, $\delta$ = 0.1, $\lambda_s$ = 1, $\sigma$ = 3 cm$^{-1}$.}
	\label{figS7}
\end{figure}

\begin{figure}[ht!]
	\centering
	\includegraphics[width=6.5cm]{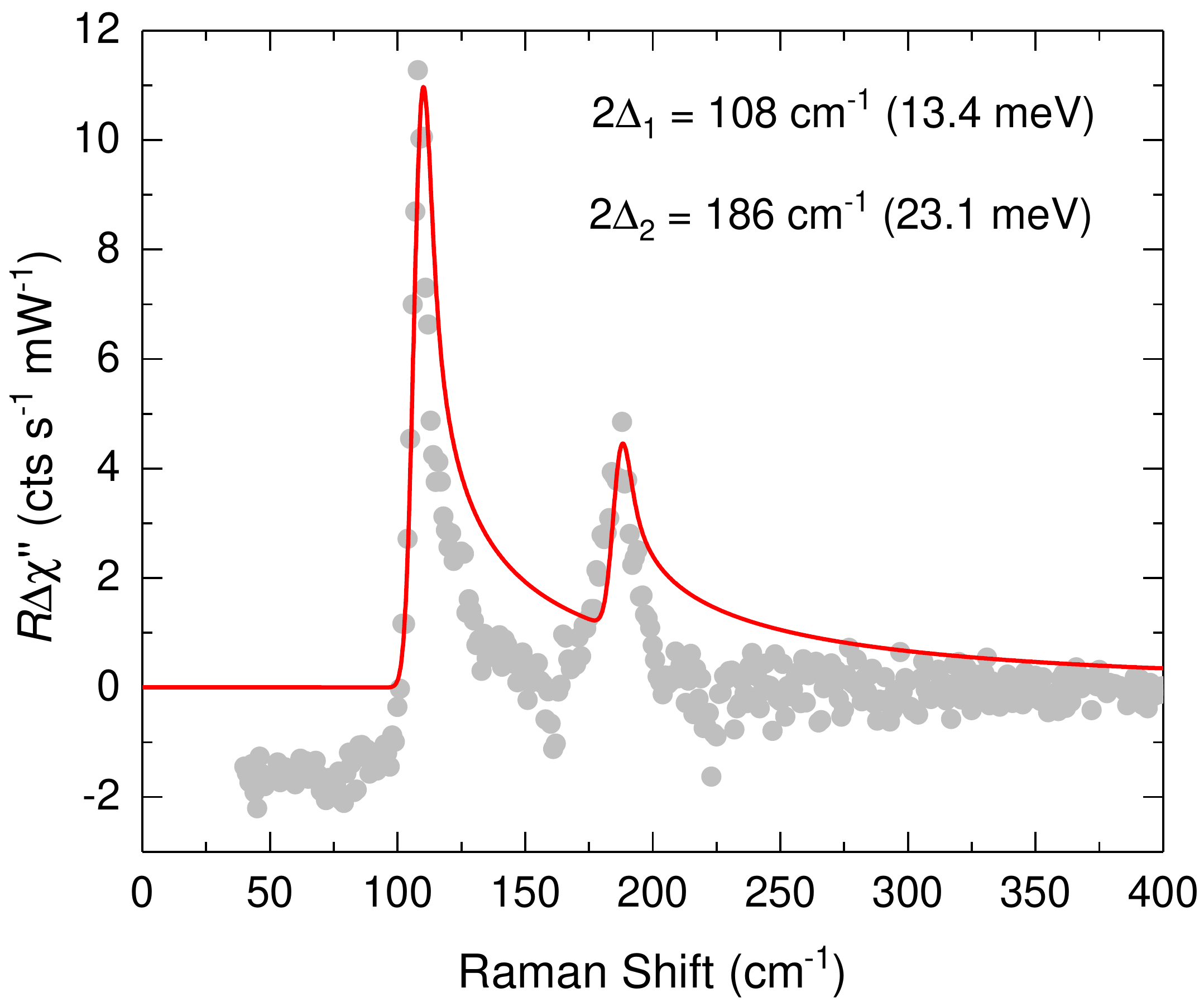}
  \caption{\rev{Difference Raman spectrum between 7.2\,K and 46\,K in \Blg symmetry (grey points) and fitting with two istropic s-wave gaps (red curve). }}
	\label{figS8}
\end{figure}

\noindent \textbf{G. Phenomenology of collective modes}
~\\

For describing the data we adopted the phenomenology proposed by Scalapino and Devereaux \cite{Scalapino:2009}. In this theory, the $s$-wave state pairing strength exceeds that of the $d$-wave state. Therefore, except for the $s$-wave ground state, there could be a $d$-wave (\Blg) collective excitonic mode. The Raman scattering is given by

\begin{widetext}
\begin{eqnarray}
R\chi^{''}(\omega)=\frac{4\pi}{\omega}\left\langle\frac{\Delta(\theta)^2}{\sqrt{\omega^2-(2\Delta(\theta))^2}}\right\rangle+Im\left\{\frac{2\left\langle\frac{2\Delta(\theta)}{\omega}\overline{P}(\omega,\theta)\right\rangle^2}{\left(\frac{1}{\lambda_d}-\frac{1}{\widetilde{\lambda_s}}\right)-\langle\overline{P}(\omega,\theta)\rangle-i\delta}\right\}.
  \label{eqn:12}
\end{eqnarray}
\end{widetext}

Here, $\Delta(\theta)$ is the $s$-wave ground-state gap. $\lambda_s$ and $\lambda_d$ are the $s$-wave and $d$-wave pairing strength respectively. A small damping term $\delta$ has been added to control the integrals. The function $\overline{P}(\omega,\theta)$ is defined by

\begin{widetext}
\begin{eqnarray}
\overline{P}(\omega,\theta)=\left\{
\begin{aligned}
&\frac{\omega/2\Delta(\theta)}{\sqrt{1-[\omega/2\Delta(\theta)]^2}}\sin^{-1}\left[\frac{\omega}{2\Delta(\theta)}\right], \left|\frac{\omega}{2\Delta(\theta)}\right|<1\\
&\frac{\omega/2\Delta(\theta)}{\sqrt{[\omega/2\Delta(\theta)]^2-1}}\left[\ln\left\{\left|\frac{\omega}{2\Delta(\theta)}\right|-\sqrt{[\omega/2\Delta(\theta)]^2-1}\right\}+i\frac{\pi}{2}\right], \left|\frac{\omega}{2\Delta(\theta)}\right|>1\\
\end{aligned}
\right.
  \label{eqn:8}
\end{eqnarray}
\end{widetext}
~\\

Since the line at 110\,cm$^{-1}$  is very narrow the resolution of the spectrometer cannot be neglected and a convolution with a Gaussian is necessary for properly describing the Raman scattering intensity
\begin{equation}
I(\omega_0)=\frac{R}{\sqrt{2\pi}\sigma}\int\chi''{(\omega)}e^{\frac{(\omega-\omega_0)^2}{2\sigma^2}}d\omega,
  \label{eqn:13}
\end{equation}

\noindent where $\sigma^2$ is the variance of the Gaussian.

Figure~\ref{figS7} shows plots of the Raman response for different coupling strength $\lambda_d$. With the increase of the $\lambda_d$, the pair breaking peak gets suppressed. In our case, there are two distinct superconducting energy gaps. To simplify, we treat them as two independent isotropic gaps $\Delta_i$ on the different bands as suggested by the tunneling experiments, that contribute to Raman scattering separately. The total Raman intensity is expressed as
\begin{equation}
I_{\rm total}(\omega)= \sum_{i=1,2}b_iI_i(\omega),
  \label{eqn:13}
\end{equation}

\noindent where $b_i$ is a dimensionless constant.

\rev{Based on this phenomenological model, we fit the difference Raman spectrum in \Blg symmetry by considering three situations, i.e. two pair-breaking peaks (see Fig.~\ref{figS7}), one BS or Leggett mode plus one pairing-breaking mode (see Fig.4(b) in the main text) and two BS modes (see Fig.4(d) in the main text). The fitting parameters are listed in Table~\ref{tab1}. It is obvious that the two peaks in \Blg spectrum can not be explained by only considering two isotropic s-wave gaps.}


\nocite{*}
\bibliography{refs} 

\begin{thebibliography}{48}%
\makeatletter
\providecommand \@ifxundefined [1]{%
 \@ifx{#1\undefined}
}%
\providecommand \@ifnum [1]{%
 \ifnum #1\expandafter \@firstoftwo
 \else \expandafter \@secondoftwo
 \fi
}%
\providecommand \@ifx [1]{%
 \ifx #1\expandafter \@firstoftwo
 \else \expandafter \@secondoftwo
 \fi
}%
\providecommand \natexlab [1]{#1}%
\providecommand \enquote  [1]{``#1''}%
\providecommand \bibnamefont  [1]{#1}%
\providecommand \bibfnamefont [1]{#1}%
\providecommand \citenamefont [1]{#1}%
\providecommand \href@noop [0]{\@secondoftwo}%
\providecommand \href [0]{\begingroup \@sanitize@url \@href}%
\providecommand \@href[1]{\@@startlink{#1}\@@href}%
\providecommand \@@href[1]{\endgroup#1\@@endlink}%
\providecommand \@sanitize@url [0]{\catcode `\\12\catcode `\$12\catcode
  `\&12\catcode `\#12\catcode `\^12\catcode `\_12\catcode `\%12\relax}%
\providecommand \@@startlink[1]{}%
\providecommand \@@endlink[0]{}%
\providecommand \url  [0]{\begingroup\@sanitize@url \@url }%
\providecommand \@url [1]{\endgroup\@href {#1}{\urlprefix }}%
\providecommand \urlprefix  [0]{URL }%
\providecommand \Eprint [0]{\href }%
\providecommand \doibase [0]{http://dx.doi.org/}%
\providecommand \selectlanguage [0]{\@gobble}%
\providecommand \bibinfo  [0]{\@secondoftwo}%
\providecommand \bibfield  [0]{\@secondoftwo}%
\providecommand \translation [1]{[#1]}%
\providecommand \BibitemOpen [0]{}%
\providecommand \bibitemStop [0]{}%
\providecommand \bibitemNoStop [0]{.\EOS\space}%
\providecommand \EOS [0]{\spacefactor3000\relax}%
\providecommand \BibitemShut  [1]{\csname bibitem#1\endcsname}%
\let\auto@bib@innerbib\@empty
\bibitem [{\citenamefont {Paglione}\ and\ \citenamefont
  {Greene}(2010)}]{Paglione:2010}%
  \BibitemOpen
  \bibfield  {author} {\bibinfo {author} {\bibfnamefont {J.}~\bibnamefont
  {Paglione}}\ and\ \bibinfo {author} {\bibfnamefont {R.~L.}\ \bibnamefont
  {Greene}},\ }\href {\doibase 10.1038/Nphys1759} {\bibfield  {journal}
  {\bibinfo  {journal} {Nature Phys.}\ }\textbf {\bibinfo {volume} {6}},\
  \bibinfo {pages} {645} (\bibinfo {year} {2010})}\BibitemShut {NoStop}%
\bibitem [{\citenamefont {Mazin}\ \emph {et~al.}(2008)\citenamefont {Mazin},
  \citenamefont {Singh}, \citenamefont {Johannes},\ and\ \citenamefont
  {Du}}]{Mazin:2008}%
  \BibitemOpen
  \bibfield  {author} {\bibinfo {author} {\bibfnamefont {I.~I.}\ \bibnamefont
  {Mazin}}, \bibinfo {author} {\bibfnamefont {D.~J.}\ \bibnamefont {Singh}},
  \bibinfo {author} {\bibfnamefont {M.~D.}\ \bibnamefont {Johannes}}, \ and\
  \bibinfo {author} {\bibfnamefont {M.~H.}\ \bibnamefont {Du}},\ }\href
  {\doibase 10.1103/Physrevlett.101.057003} {\bibfield  {journal} {\bibinfo
  {journal} {Phys. Rev. Lett.}\ }\textbf {\bibinfo {volume} {101}},\ \bibinfo
  {pages} {057003} (\bibinfo {year} {2008})}\BibitemShut {NoStop}%
\bibitem [{\citenamefont {Scalapino}(2012)}]{Scalapino:2012}%
  \BibitemOpen
  \bibfield  {author} {\bibinfo {author} {\bibfnamefont {D.~J.}\ \bibnamefont
  {Scalapino}},\ }\href {\doibase 10.1103/RevModPhys.84.1383} {\bibfield
  {journal} {\bibinfo  {journal} {Rev. Mod. Phys.}\ }\textbf {\bibinfo {volume}
  {84}},\ \bibinfo {pages} {1383} (\bibinfo {year} {2012})}\BibitemShut
  {NoStop}%
\bibitem [{\citenamefont {Onari}\ and\ \citenamefont
  {Kontani}(2009)}]{Onari:2009}%
  \BibitemOpen
  \bibfield  {author} {\bibinfo {author} {\bibfnamefont {S.}~\bibnamefont
  {Onari}}\ and\ \bibinfo {author} {\bibfnamefont {H.}~\bibnamefont
  {Kontani}},\ }\href {\doibase 10.1103/Physrevlett.103.177001} {\bibfield
  {journal} {\bibinfo  {journal} {Phys. Rev. Lett.}\ }\textbf {\bibinfo
  {volume} {103}},\ \bibinfo {pages} {177001} (\bibinfo {year}
  {2009})}\BibitemShut {NoStop}%
\bibitem [{\citenamefont {Kontani}\ and\ \citenamefont
  {Onari}(2010)}]{Kontani:2010}%
  \BibitemOpen
  \bibfield  {author} {\bibinfo {author} {\bibfnamefont {H.}~\bibnamefont
  {Kontani}}\ and\ \bibinfo {author} {\bibfnamefont {S.}~\bibnamefont
  {Onari}},\ }\href {\doibase 10.1103/Physrevlett.104.157001} {\bibfield
  {journal} {\bibinfo  {journal} {Phys. Rev. Lett.}\ }\textbf {\bibinfo
  {volume} {104}},\ \bibinfo {pages} {157001} (\bibinfo {year}
  {2010})}\BibitemShut {NoStop}%
\bibitem [{\citenamefont {Borisenko}\ \emph {et~al.}(2016)\citenamefont
  {Borisenko}, \citenamefont {Evtushinsky}, \citenamefont {Liu}, \citenamefont
  {Morozov}, \citenamefont {Kappenberger}, \citenamefont {Wurmehl},
  \citenamefont {Buchner}, \citenamefont {Yaresko}, \citenamefont {Kim},
  \citenamefont {Hoesch}, \citenamefont {Wolf},\ and\ \citenamefont
  {Zhigadlo}}]{Borisenko:2016}%
  \BibitemOpen
  \bibfield  {author} {\bibinfo {author} {\bibfnamefont {S.~V.}\ \bibnamefont
  {Borisenko}}, \bibinfo {author} {\bibfnamefont {D.~V.}\ \bibnamefont
  {Evtushinsky}}, \bibinfo {author} {\bibfnamefont {Z.~H.}\ \bibnamefont
  {Liu}}, \bibinfo {author} {\bibfnamefont {I.}~\bibnamefont {Morozov}},
  \bibinfo {author} {\bibfnamefont {R.}~\bibnamefont {Kappenberger}}, \bibinfo
  {author} {\bibfnamefont {S.}~\bibnamefont {Wurmehl}}, \bibinfo {author}
  {\bibfnamefont {B.}~\bibnamefont {Buchner}}, \bibinfo {author} {\bibfnamefont
  {A.~N.}\ \bibnamefont {Yaresko}}, \bibinfo {author} {\bibfnamefont {T.~K.}\
  \bibnamefont {Kim}}, \bibinfo {author} {\bibfnamefont {M.}~\bibnamefont
  {Hoesch}}, \bibinfo {author} {\bibfnamefont {T.}~\bibnamefont {Wolf}}, \ and\
  \bibinfo {author} {\bibfnamefont {N.~D.}\ \bibnamefont {Zhigadlo}},\ }\href
  {\doibase 10.1038/Nphys3594} {\bibfield  {journal} {\bibinfo  {journal}
  {Nature Phys.}\ }\textbf {\bibinfo {volume} {12}},\ \bibinfo {pages} {311}
  (\bibinfo {year} {2016})}\BibitemShut {NoStop}%
\bibitem [{\citenamefont {Lederer}\ \emph {et~al.}(2015)\citenamefont
  {Lederer}, \citenamefont {Schattner}, \citenamefont {Berg},\ and\
  \citenamefont {Kivelson}}]{Lederer:2015}%
  \BibitemOpen
  \bibfield  {author} {\bibinfo {author} {\bibfnamefont {S.}~\bibnamefont
  {Lederer}}, \bibinfo {author} {\bibfnamefont {Y.}~\bibnamefont {Schattner}},
  \bibinfo {author} {\bibfnamefont {E.}~\bibnamefont {Berg}}, \ and\ \bibinfo
  {author} {\bibfnamefont {S.~A.}\ \bibnamefont {Kivelson}},\ }\href {\doibase
  10.1103/Physrevlett.114.097001} {\bibfield  {journal} {\bibinfo  {journal}
  {Phys. Rev. Lett.}\ }\textbf {\bibinfo {volume} {114}},\ \bibinfo {pages}
  {097001} (\bibinfo {year} {2015})}\BibitemShut {NoStop}%
\bibitem [{\citenamefont {Lu}\ \emph {et~al.}(2015)\citenamefont {Lu},
  \citenamefont {Wang}, \citenamefont {Wu}, \citenamefont {Wu}, \citenamefont
  {Zhao}, \citenamefont {Zeng}, \citenamefont {Luo}, \citenamefont {Wu},
  \citenamefont {Bao}, \citenamefont {Zhang}, \citenamefont {Huang},
  \citenamefont {Huang},\ and\ \citenamefont {Chen}}]{Lu:2015}%
  \BibitemOpen
  \bibfield  {author} {\bibinfo {author} {\bibfnamefont {X.~F.}\ \bibnamefont
  {Lu}}, \bibinfo {author} {\bibfnamefont {N.~Z.}\ \bibnamefont {Wang}},
  \bibinfo {author} {\bibfnamefont {H.}~\bibnamefont {Wu}}, \bibinfo {author}
  {\bibfnamefont {Y.~P.}\ \bibnamefont {Wu}}, \bibinfo {author} {\bibfnamefont
  {D.}~\bibnamefont {Zhao}}, \bibinfo {author} {\bibfnamefont {X.~Z.}\
  \bibnamefont {Zeng}}, \bibinfo {author} {\bibfnamefont {X.~G.}\ \bibnamefont
  {Luo}}, \bibinfo {author} {\bibfnamefont {T.}~\bibnamefont {Wu}}, \bibinfo
  {author} {\bibfnamefont {W.}~\bibnamefont {Bao}}, \bibinfo {author}
  {\bibfnamefont {G.~H.}\ \bibnamefont {Zhang}}, \bibinfo {author}
  {\bibfnamefont {F.~Q.}\ \bibnamefont {Huang}}, \bibinfo {author}
  {\bibfnamefont {Q.~Z.}\ \bibnamefont {Huang}}, \ and\ \bibinfo {author}
  {\bibfnamefont {X.~H.}\ \bibnamefont {Chen}},\ }\href {\doibase
  10.1038/nmat4155} {\bibfield  {journal} {\bibinfo  {journal} {Nature Mater.}\
  }\textbf {\bibinfo {volume} {14}},\ \bibinfo {pages} {325} (\bibinfo {year}
  {2015})}\BibitemShut {NoStop}%
\bibitem [{\citenamefont {Burrard-Lucas}\ \emph {et~al.}(2013)\citenamefont
  {Burrard-Lucas}, \citenamefont {Free}, \citenamefont {Sedlmaier},
  \citenamefont {Wright}, \citenamefont {Cassidy}, \citenamefont {Hara},
  \citenamefont {Corkett}, \citenamefont {Lancaster}, \citenamefont {Baker},
  \citenamefont {Blundell},\ and\ \citenamefont {Clarke}}]{Burrard-Lucas:2013}%
  \BibitemOpen
  \bibfield  {author} {\bibinfo {author} {\bibfnamefont {M.}~\bibnamefont
  {Burrard-Lucas}}, \bibinfo {author} {\bibfnamefont {D.~G.}\ \bibnamefont
  {Free}}, \bibinfo {author} {\bibfnamefont {S.~J.}\ \bibnamefont {Sedlmaier}},
  \bibinfo {author} {\bibfnamefont {J.~D.}\ \bibnamefont {Wright}}, \bibinfo
  {author} {\bibfnamefont {S.~J.}\ \bibnamefont {Cassidy}}, \bibinfo {author}
  {\bibfnamefont {Y.}~\bibnamefont {Hara}}, \bibinfo {author} {\bibfnamefont
  {A.~J.}\ \bibnamefont {Corkett}}, \bibinfo {author} {\bibfnamefont
  {T.}~\bibnamefont {Lancaster}}, \bibinfo {author} {\bibfnamefont {P.~J.}\
  \bibnamefont {Baker}}, \bibinfo {author} {\bibfnamefont {S.~J.}\ \bibnamefont
  {Blundell}}, \ and\ \bibinfo {author} {\bibfnamefont {S.~J.}\ \bibnamefont
  {Clarke}},\ }\href {\doibase 10.1038/NMAT3464} {\bibfield  {journal}
  {\bibinfo  {journal} {Nature Mater.}\ }\textbf {\bibinfo {volume} {12}},\
  \bibinfo {pages} {15} (\bibinfo {year} {2013})}\BibitemShut {NoStop}%
\bibitem [{\citenamefont {Zhao}\ \emph {et~al.}(2016)\citenamefont {Zhao},
  \citenamefont {Liang}, \citenamefont {Yuan}, \citenamefont {Hu},
  \citenamefont {Liu}, \citenamefont {Huang}, \citenamefont {He}, \citenamefont
  {Shen}, \citenamefont {Xu}, \citenamefont {Liu}, \citenamefont {Yu},
  \citenamefont {Liu}, \citenamefont {Zhou}, \citenamefont {Huang},
  \citenamefont {Dong}, \citenamefont {Zhou}, \citenamefont {Liu},
  \citenamefont {Lu}, \citenamefont {Zhao}, \citenamefont {Chen}, \citenamefont
  {Xu},\ and\ \citenamefont {Zhou}}]{Zhao:2016}%
  \BibitemOpen
  \bibfield  {author} {\bibinfo {author} {\bibfnamefont {L.}~\bibnamefont
  {Zhao}}, \bibinfo {author} {\bibfnamefont {A.~J.}\ \bibnamefont {Liang}},
  \bibinfo {author} {\bibfnamefont {D.~N.}\ \bibnamefont {Yuan}}, \bibinfo
  {author} {\bibfnamefont {Y.}~\bibnamefont {Hu}}, \bibinfo {author}
  {\bibfnamefont {D.~F.}\ \bibnamefont {Liu}}, \bibinfo {author} {\bibfnamefont
  {J.~W.}\ \bibnamefont {Huang}}, \bibinfo {author} {\bibfnamefont {S.~L.}\
  \bibnamefont {He}}, \bibinfo {author} {\bibfnamefont {B.}~\bibnamefont
  {Shen}}, \bibinfo {author} {\bibfnamefont {Y.}~\bibnamefont {Xu}}, \bibinfo
  {author} {\bibfnamefont {X.}~\bibnamefont {Liu}}, \bibinfo {author}
  {\bibfnamefont {L.}~\bibnamefont {Yu}}, \bibinfo {author} {\bibfnamefont
  {G.~D.}\ \bibnamefont {Liu}}, \bibinfo {author} {\bibfnamefont {H.~X.}\
  \bibnamefont {Zhou}}, \bibinfo {author} {\bibfnamefont {Y.~L.}\ \bibnamefont
  {Huang}}, \bibinfo {author} {\bibfnamefont {X.~L.}\ \bibnamefont {Dong}},
  \bibinfo {author} {\bibfnamefont {F.}~\bibnamefont {Zhou}}, \bibinfo {author}
  {\bibfnamefont {K.}~\bibnamefont {Liu}}, \bibinfo {author} {\bibfnamefont
  {Z.~Y.}\ \bibnamefont {Lu}}, \bibinfo {author} {\bibfnamefont {Z.~X.}\
  \bibnamefont {Zhao}}, \bibinfo {author} {\bibfnamefont {C.~T.}\ \bibnamefont
  {Chen}}, \bibinfo {author} {\bibfnamefont {Z.~Y.}\ \bibnamefont {Xu}}, \ and\
  \bibinfo {author} {\bibfnamefont {X.~J.}\ \bibnamefont {Zhou}},\ }\href
  {\doibase 10.1038/Ncomms10608} {\bibfield  {journal} {\bibinfo  {journal}
  {Nature Communi.}\ }\textbf {\bibinfo {volume} {7}},\ \bibinfo {pages}
  {10608} (\bibinfo {year} {2016})}\BibitemShut {NoStop}%
\bibitem [{\citenamefont {Nekrasov}\ and\ \citenamefont
  {Sadovskii}(2015)}]{Nekrasov:2015}%
  \BibitemOpen
  \bibfield  {author} {\bibinfo {author} {\bibfnamefont {I.~A.}\ \bibnamefont
  {Nekrasov}}\ and\ \bibinfo {author} {\bibfnamefont {M.~V.}\ \bibnamefont
  {Sadovskii}},\ }\href {\doibase 10.1134/S0021364015010105} {\bibfield
  {journal} {\bibinfo  {journal} {JETP Lett.}\ }\textbf {\bibinfo {volume}
  {101}},\ \bibinfo {pages} {47} (\bibinfo {year} {2015})}\BibitemShut
  {NoStop}%
\bibitem [{\citenamefont {Shi}\ \emph {et~al.}(2017)\citenamefont {Shi},
  \citenamefont {Han}, \citenamefont {Peng}, \citenamefont {Richard},
  \citenamefont {Qian}, \citenamefont {Wu}, \citenamefont {Qiu}, \citenamefont
  {Wang}, \citenamefont {Hu}, \citenamefont {Sun},\ and\ \citenamefont
  {Ding}}]{Shi:2017}%
  \BibitemOpen
  \bibfield  {author} {\bibinfo {author} {\bibfnamefont {X.}~\bibnamefont
  {Shi}}, \bibinfo {author} {\bibfnamefont {Z.-Q.}\ \bibnamefont {Han}},
  \bibinfo {author} {\bibfnamefont {X.-L.}\ \bibnamefont {Peng}}, \bibinfo
  {author} {\bibfnamefont {P.}~\bibnamefont {Richard}}, \bibinfo {author}
  {\bibfnamefont {T.}~\bibnamefont {Qian}}, \bibinfo {author} {\bibfnamefont
  {X.-X.}\ \bibnamefont {Wu}}, \bibinfo {author} {\bibfnamefont {M.-W.}\
  \bibnamefont {Qiu}}, \bibinfo {author} {\bibfnamefont {S.~C.}\ \bibnamefont
  {Wang}}, \bibinfo {author} {\bibfnamefont {J.~P.}\ \bibnamefont {Hu}},
  \bibinfo {author} {\bibfnamefont {Y.-J.}\ \bibnamefont {Sun}}, \ and\
  \bibinfo {author} {\bibfnamefont {H.}~\bibnamefont {Ding}},\ }\href {\doibase
  10.1038/ncomms14988} {\bibfield  {journal} {\bibinfo  {journal} {Nature
  Commun.}\ }\textbf {\bibinfo {volume} {8}},\ \bibinfo {pages} {14988}
  (\bibinfo {year} {2017})}\BibitemShut {NoStop}%
\bibitem [{\citenamefont {Du}\ \emph {et~al.}(2016)\citenamefont {Du},
  \citenamefont {Yang}, \citenamefont {Lin}, \citenamefont {Fang},
  \citenamefont {Du}, \citenamefont {Xing}, \citenamefont {Yang}, \citenamefont
  {Zhu},\ and\ \citenamefont {Wen}}]{Du:2016}%
  \BibitemOpen
  \bibfield  {author} {\bibinfo {author} {\bibfnamefont {Z.~Y.}\ \bibnamefont
  {Du}}, \bibinfo {author} {\bibfnamefont {X.}~\bibnamefont {Yang}}, \bibinfo
  {author} {\bibfnamefont {H.}~\bibnamefont {Lin}}, \bibinfo {author}
  {\bibfnamefont {D.~L.}\ \bibnamefont {Fang}}, \bibinfo {author}
  {\bibfnamefont {G.}~\bibnamefont {Du}}, \bibinfo {author} {\bibfnamefont
  {J.}~\bibnamefont {Xing}}, \bibinfo {author} {\bibfnamefont {H.}~\bibnamefont
  {Yang}}, \bibinfo {author} {\bibfnamefont {X.~Y.}\ \bibnamefont {Zhu}}, \
  and\ \bibinfo {author} {\bibfnamefont {H.~H.}\ \bibnamefont {Wen}},\ }\href
  {\doibase 10.1038/Ncomms10565} {\bibfield  {journal} {\bibinfo  {journal}
  {Nature Communi.}\ }\textbf {\bibinfo {volume} {7}},\ \bibinfo {pages}
  {10565} (\bibinfo {year} {2016})}\BibitemShut {NoStop}%
\bibitem [{\citenamefont {Du}\ \emph {et~al.}(2018)\citenamefont {Du},
  \citenamefont {Yang}, \citenamefont {Altenfeld}, \citenamefont {Gu},
  \citenamefont {Yang}, \citenamefont {Eremin}, \citenamefont {Hirschfeld},
  \citenamefont {Mazin}, \citenamefont {Lin}, \citenamefont {Zhu},\ and\
  \citenamefont {Wen}}]{Du:2018}%
  \BibitemOpen
  \bibfield  {author} {\bibinfo {author} {\bibfnamefont {Z.~Y.}\ \bibnamefont
  {Du}}, \bibinfo {author} {\bibfnamefont {X.}~\bibnamefont {Yang}}, \bibinfo
  {author} {\bibfnamefont {D.}~\bibnamefont {Altenfeld}}, \bibinfo {author}
  {\bibfnamefont {Q.~Q.}\ \bibnamefont {Gu}}, \bibinfo {author} {\bibfnamefont
  {H.}~\bibnamefont {Yang}}, \bibinfo {author} {\bibfnamefont {I.}~\bibnamefont
  {Eremin}}, \bibinfo {author} {\bibfnamefont {P.~J.}\ \bibnamefont
  {Hirschfeld}}, \bibinfo {author} {\bibfnamefont {I.~I.}\ \bibnamefont
  {Mazin}}, \bibinfo {author} {\bibfnamefont {H.}~\bibnamefont {Lin}}, \bibinfo
  {author} {\bibfnamefont {X.~Y.}\ \bibnamefont {Zhu}}, \ and\ \bibinfo
  {author} {\bibfnamefont {H.~H.}\ \bibnamefont {Wen}},\ }\href {\doibase
  10.1038/Nphys4299} {\bibfield  {journal} {\bibinfo  {journal} {Nature Phys.}\
  }\textbf {\bibinfo {volume} {14}},\ \bibinfo {pages} {134} (\bibinfo {year}
  {2018})}\BibitemShut {NoStop}%
\bibitem [{\citenamefont {Chen}\ \emph {et~al.}(2019)\citenamefont {Chen},
  \citenamefont {Liu}, \citenamefont {Zhang}, \citenamefont {Li}, \citenamefont
  {Shen}, \citenamefont {Dong}, \citenamefont {Zhao}, \citenamefont {Zhang},\
  and\ \citenamefont {Feng}}]{Chen:2019}%
  \BibitemOpen
  \bibfield  {author} {\bibinfo {author} {\bibfnamefont {C.}~\bibnamefont
  {Chen}}, \bibinfo {author} {\bibfnamefont {Q.}~\bibnamefont {Liu}}, \bibinfo
  {author} {\bibfnamefont {T.~Z.}\ \bibnamefont {Zhang}}, \bibinfo {author}
  {\bibfnamefont {D.}~\bibnamefont {Li}}, \bibinfo {author} {\bibfnamefont
  {P.~P.}\ \bibnamefont {Shen}}, \bibinfo {author} {\bibfnamefont {X.~L.}\
  \bibnamefont {Dong}}, \bibinfo {author} {\bibfnamefont {Z.~X.}\ \bibnamefont
  {Zhao}}, \bibinfo {author} {\bibfnamefont {T.}~\bibnamefont {Zhang}}, \ and\
  \bibinfo {author} {\bibfnamefont {D.~L.}\ \bibnamefont {Feng}},\ }\href
  {\doibase 10.1088/0256-307x/36/5/057403} {\bibfield  {journal} {\bibinfo
  {journal} {Chin. Rev. Lett.}\ }\textbf {\bibinfo {volume} {36}},\ \bibinfo
  {pages} {057403} (\bibinfo {year} {2019})}\BibitemShut {NoStop}%
\bibitem [{\citenamefont {Devereaux}\ and\ \citenamefont
  {Hackl}(2007)}]{Devereaux:2007}%
  \BibitemOpen
  \bibfield  {author} {\bibinfo {author} {\bibfnamefont {T.~P.}\ \bibnamefont
  {Devereaux}}\ and\ \bibinfo {author} {\bibfnamefont {R.}~\bibnamefont
  {Hackl}},\ }\href {\doibase 10.1103/RevModPhys.79.175} {\bibfield  {journal}
  {\bibinfo  {journal} {Rev. Mod. Phys.}\ }\textbf {\bibinfo {volume} {79}},\
  \bibinfo {pages} {175} (\bibinfo {year} {2007})}\BibitemShut {NoStop}%
\bibitem [{\citenamefont {Scalapino}\ and\ \citenamefont
  {Devereaux}(2009)}]{Scalapino:2009}%
  \BibitemOpen
  \bibfield  {author} {\bibinfo {author} {\bibfnamefont {D.~J.}\ \bibnamefont
  {Scalapino}}\ and\ \bibinfo {author} {\bibfnamefont {T.~P.}\ \bibnamefont
  {Devereaux}},\ }\href {\doibase 10.1103/Physrevb.80.140512} {\bibfield
  {journal} {\bibinfo  {journal} {Phys. Rev. B}\ }\textbf {\bibinfo {volume}
  {80}},\ \bibinfo {pages} {140512(R)} (\bibinfo {year} {2009})}\BibitemShut
  {NoStop}%
\bibitem [{\citenamefont {Maiti}\ \emph {et~al.}(2016)\citenamefont {Maiti},
  \citenamefont {Maier}, \citenamefont {B\"ohm}, \citenamefont {Hackl},\ and\
  \citenamefont {Hirschfeld}}]{Maiti:2016}%
  \BibitemOpen
  \bibfield  {author} {\bibinfo {author} {\bibfnamefont {S.}~\bibnamefont
  {Maiti}}, \bibinfo {author} {\bibfnamefont {T.~A.}\ \bibnamefont {Maier}},
  \bibinfo {author} {\bibfnamefont {T.}~\bibnamefont {B\"ohm}}, \bibinfo
  {author} {\bibfnamefont {R.}~\bibnamefont {Hackl}}, \ and\ \bibinfo {author}
  {\bibfnamefont {P.~J.}\ \bibnamefont {Hirschfeld}},\ }\href {\doibase
  10.1103/Physrevlett.117.257001} {\bibfield  {journal} {\bibinfo  {journal}
  {Phys. Rev. Lett.}\ }\textbf {\bibinfo {volume} {117}},\ \bibinfo {pages}
  {257001} (\bibinfo {year} {2016})}\BibitemShut {NoStop}%
\bibitem [{\citenamefont {Kretzschmar}\ \emph {et~al.}(2013)\citenamefont
  {Kretzschmar}, \citenamefont {Muschler}, \citenamefont {B\"ohm},
  \citenamefont {Baum}, \citenamefont {Hackl}, \citenamefont {Wen},
  \citenamefont {Tsurkan}, \citenamefont {Deisenhofer},\ and\ \citenamefont
  {Loidl}}]{Kretzschmar:2013}%
  \BibitemOpen
  \bibfield  {author} {\bibinfo {author} {\bibfnamefont {F.}~\bibnamefont
  {Kretzschmar}}, \bibinfo {author} {\bibfnamefont {B.}~\bibnamefont
  {Muschler}}, \bibinfo {author} {\bibfnamefont {T.}~\bibnamefont {B\"ohm}},
  \bibinfo {author} {\bibfnamefont {A.}~\bibnamefont {Baum}}, \bibinfo {author}
  {\bibfnamefont {R.}~\bibnamefont {Hackl}}, \bibinfo {author} {\bibfnamefont
  {H.~H.}\ \bibnamefont {Wen}}, \bibinfo {author} {\bibfnamefont
  {V.}~\bibnamefont {Tsurkan}}, \bibinfo {author} {\bibfnamefont
  {J.}~\bibnamefont {Deisenhofer}}, \ and\ \bibinfo {author} {\bibfnamefont
  {A.}~\bibnamefont {Loidl}},\ }\href {\doibase 10.1103/Physrevlett.110.187002}
  {\bibfield  {journal} {\bibinfo  {journal} {Phys. Rev. Lett.}\ }\textbf
  {\bibinfo {volume} {110}},\ \bibinfo {pages} {187002} (\bibinfo {year}
  {2013})}\BibitemShut {NoStop}%
\bibitem [{\citenamefont {B\"ohm}\ \emph {et~al.}(2014)\citenamefont {B\"ohm},
  \citenamefont {Kemper}, \citenamefont {Moritz}, \citenamefont {Kretzschmar},
  \citenamefont {Muschler}, \citenamefont {Eiter}, \citenamefont {Hackl},
  \citenamefont {Devereaux}, \citenamefont {Scalapino},\ and\ \citenamefont
  {Wen}}]{Bohm:2014}%
  \BibitemOpen
  \bibfield  {author} {\bibinfo {author} {\bibfnamefont {T.}~\bibnamefont
  {B\"ohm}}, \bibinfo {author} {\bibfnamefont {A.~F.}\ \bibnamefont {Kemper}},
  \bibinfo {author} {\bibfnamefont {B.}~\bibnamefont {Moritz}}, \bibinfo
  {author} {\bibfnamefont {F.}~\bibnamefont {Kretzschmar}}, \bibinfo {author}
  {\bibfnamefont {B.}~\bibnamefont {Muschler}}, \bibinfo {author}
  {\bibfnamefont {H.~M.}\ \bibnamefont {Eiter}}, \bibinfo {author}
  {\bibfnamefont {R.}~\bibnamefont {Hackl}}, \bibinfo {author} {\bibfnamefont
  {T.~P.}\ \bibnamefont {Devereaux}}, \bibinfo {author} {\bibfnamefont {D.~J.}\
  \bibnamefont {Scalapino}}, \ and\ \bibinfo {author} {\bibfnamefont {H.~H.}\
  \bibnamefont {Wen}},\ }\href {\doibase 104610.1103/Physrevx.4.041046}
  {\bibfield  {journal} {\bibinfo  {journal} {Phys. Rev. X}\ }\textbf {\bibinfo
  {volume} {4}},\ \bibinfo {pages} {041046} (\bibinfo {year}
  {2014})}\BibitemShut {NoStop}%
\bibitem [{\citenamefont {Abrikosov}\ and\ \citenamefont
  {Fal'kovskii}(1961)}]{Abrikosov:1961}%
  \BibitemOpen
  \bibfield  {author} {\bibinfo {author} {\bibfnamefont {A.~A.}\ \bibnamefont
  {Abrikosov}}\ and\ \bibinfo {author} {\bibfnamefont {L.~A.}\ \bibnamefont
  {Fal'kovskii}},\ }\href@noop {} {\bibfield  {journal} {\bibinfo  {journal}
  {Zh. Eksp. Teor. Fiz.}\ }\textbf {\bibinfo {volume} {40}},\ \bibinfo {pages}
  {262} (\bibinfo {year} {1961})},\ \bibinfo {note} {[Sov. Phys. JETP {\bf 13},
  179 (1961)]}\BibitemShut {NoStop}%
\bibitem [{\citenamefont {Klein}\ and\ \citenamefont
  {Dierker}(1984)}]{Klein:1984}%
  \BibitemOpen
  \bibfield  {author} {\bibinfo {author} {\bibfnamefont {M.~V.}\ \bibnamefont
  {Klein}}\ and\ \bibinfo {author} {\bibfnamefont {S.~B.}\ \bibnamefont
  {Dierker}},\ }\href {\doibase 10.1103/PhysRevB.29.4976} {\bibfield  {journal}
  {\bibinfo  {journal} {Phys. Rev. B}\ }\textbf {\bibinfo {volume} {29}},\
  \bibinfo {pages} {4976} (\bibinfo {year} {1984})}\BibitemShut {NoStop}%
\bibitem [{\citenamefont {Bardasis}\ and\ \citenamefont
  {Schrieffer}(1961)}]{Bardasis:1961}%
  \BibitemOpen
  \bibfield  {author} {\bibinfo {author} {\bibfnamefont {A.}~\bibnamefont
  {Bardasis}}\ and\ \bibinfo {author} {\bibfnamefont {J.~R.}\ \bibnamefont
  {Schrieffer}},\ }\href {\doibase 10.1103/PhysRev.121.1050} {\bibfield
  {journal} {\bibinfo  {journal} {Phys. Rev.}\ }\textbf {\bibinfo {volume}
  {121}},\ \bibinfo {pages} {1050} (\bibinfo {year} {1961})}\BibitemShut
  {NoStop}%
\bibitem [{\citenamefont {Leggett}(1966)}]{Leggett:1966}%
  \BibitemOpen
  \bibfield  {author} {\bibinfo {author} {\bibfnamefont {A.~J.}\ \bibnamefont
  {Leggett}},\ }\href {\doibase 10.1143/Ptp.36.901} {\bibfield  {journal}
  {\bibinfo  {journal} {Prog.Theor. Phys.}\ }\textbf {\bibinfo {volume} {36}},\
  \bibinfo {pages} {901} (\bibinfo {year} {1966})}\BibitemShut {NoStop}%
\bibitem [{\citenamefont {Klein}(2010)}]{Klein:2010}%
  \BibitemOpen
  \bibfield  {author} {\bibinfo {author} {\bibfnamefont {M.~V.}\ \bibnamefont
  {Klein}},\ }\href {\doibase 10.1103/Physrevb.82.014507} {\bibfield  {journal}
  {\bibinfo  {journal} {Phys. Rev. B}\ }\textbf {\bibinfo {volume} {82}},\
  \bibinfo {pages} {014507} (\bibinfo {year} {2010})}\BibitemShut {NoStop}%
\bibitem [{\citenamefont {Blumberg}\ \emph {et~al.}(2007)\citenamefont
  {Blumberg}, \citenamefont {Mialitsin}, \citenamefont {Dennis}, \citenamefont
  {Klein}, \citenamefont {Zhigadlo},\ and\ \citenamefont
  {Karpinski}}]{Blumberg:2007}%
  \BibitemOpen
  \bibfield  {author} {\bibinfo {author} {\bibfnamefont {G.}~\bibnamefont
  {Blumberg}}, \bibinfo {author} {\bibfnamefont {A.}~\bibnamefont {Mialitsin}},
  \bibinfo {author} {\bibfnamefont {B.~S.}\ \bibnamefont {Dennis}}, \bibinfo
  {author} {\bibfnamefont {M.~V.}\ \bibnamefont {Klein}}, \bibinfo {author}
  {\bibfnamefont {N.~D.}\ \bibnamefont {Zhigadlo}}, \ and\ \bibinfo {author}
  {\bibfnamefont {J.}~\bibnamefont {Karpinski}},\ }\href {\doibase
  10.1103/PhysRevLett.99.227002} {\bibfield  {journal} {\bibinfo  {journal}
  {Phys. Rev. Lett.}\ }\textbf {\bibinfo {volume} {99}},\ \bibinfo {pages}
  {227002} (\bibinfo {year} {2007})}\BibitemShut {NoStop}%
\bibitem [{\citenamefont {Huang}\ \emph
  {et~al.}(2017{\natexlab{a}})\citenamefont {Huang}, \citenamefont {Feng},
  \citenamefont {Ni}, \citenamefont {Li}, \citenamefont {Hu}, \citenamefont
  {Liu}, \citenamefont {Mao}, \citenamefont {Zhou}, \citenamefont {Zhou},
  \citenamefont {Jin}, \citenamefont {Wang}, \citenamefont {Yuan},
  \citenamefont {Dong},\ and\ \citenamefont {Zhao}}]{Huang:2017}%
  \BibitemOpen
  \bibfield  {author} {\bibinfo {author} {\bibfnamefont {Y.~L.}\ \bibnamefont
  {Huang}}, \bibinfo {author} {\bibfnamefont {Z.~P.}\ \bibnamefont {Feng}},
  \bibinfo {author} {\bibfnamefont {S.~L.}\ \bibnamefont {Ni}}, \bibinfo
  {author} {\bibfnamefont {J.}~\bibnamefont {Li}}, \bibinfo {author}
  {\bibfnamefont {W.}~\bibnamefont {Hu}}, \bibinfo {author} {\bibfnamefont
  {S.~B.}\ \bibnamefont {Liu}}, \bibinfo {author} {\bibfnamefont {Y.~Y.}\
  \bibnamefont {Mao}}, \bibinfo {author} {\bibfnamefont {H.~X.}\ \bibnamefont
  {Zhou}}, \bibinfo {author} {\bibfnamefont {F.}~\bibnamefont {Zhou}}, \bibinfo
  {author} {\bibfnamefont {K.}~\bibnamefont {Jin}}, \bibinfo {author}
  {\bibfnamefont {H.~B.}\ \bibnamefont {Wang}}, \bibinfo {author}
  {\bibfnamefont {J.}~\bibnamefont {Yuan}}, \bibinfo {author} {\bibfnamefont
  {X.~L.}\ \bibnamefont {Dong}}, \ and\ \bibinfo {author} {\bibfnamefont
  {Z.~X.}\ \bibnamefont {Zhao}},\ }\href {\doibase
  10.1088/0256-307x/34/7/077404} {\bibfield  {journal} {\bibinfo  {journal}
  {Chin. Rev. Lett.}\ }\textbf {\bibinfo {volume} {34}},\ \bibinfo {pages}
  {077404} (\bibinfo {year} {2017}{\natexlab{a}})}\BibitemShut {NoStop}%
\bibitem [{\citenamefont {Huang}\ \emph
  {et~al.}(2017{\natexlab{b}})\citenamefont {Huang}, \citenamefont {Feng},
  \citenamefont {Yuan}, \citenamefont {Hu}, \citenamefont {Li}, \citenamefont
  {Ni}, \citenamefont {Liu}, \citenamefont {Mao}, \citenamefont {Zhou},
  \citenamefont {Wang}, \citenamefont {Zhou}, \citenamefont {Zhang},
  \citenamefont {Jin}, \citenamefont {Dong},\ and\ \citenamefont
  {Zhao}}]{Huang:20172}%
  \BibitemOpen
  \bibfield  {author} {\bibinfo {author} {\bibfnamefont {Y.~L.}\ \bibnamefont
  {Huang}}, \bibinfo {author} {\bibfnamefont {Z.~P.}\ \bibnamefont {Feng}},
  \bibinfo {author} {\bibfnamefont {J.}~\bibnamefont {Yuan}}, \bibinfo {author}
  {\bibfnamefont {W.}~\bibnamefont {Hu}}, \bibinfo {author} {\bibfnamefont
  {J.}~\bibnamefont {Li}}, \bibinfo {author} {\bibfnamefont {S.~L.}\
  \bibnamefont {Ni}}, \bibinfo {author} {\bibfnamefont {S.~B.}\ \bibnamefont
  {Liu}}, \bibinfo {author} {\bibfnamefont {Y.~Y.}\ \bibnamefont {Mao}},
  \bibinfo {author} {\bibfnamefont {H.~X.}\ \bibnamefont {Zhou}}, \bibinfo
  {author} {\bibfnamefont {H.~B.}\ \bibnamefont {Wang}}, \bibinfo {author}
  {\bibfnamefont {F.}~\bibnamefont {Zhou}}, \bibinfo {author} {\bibfnamefont
  {G.~M.}\ \bibnamefont {Zhang}}, \bibinfo {author} {\bibfnamefont
  {K.}~\bibnamefont {Jin}}, \bibinfo {author} {\bibfnamefont {X.~L.}\
  \bibnamefont {Dong}}, \ and\ \bibinfo {author} {\bibfnamefont {Z.~X.}\
  \bibnamefont {Zhao}},\ }\href {https://arxiv.org/abs/1711.02920} {\bibfield
  {journal} {\bibinfo  {journal} {arXiv: 1711.02920}\ } (\bibinfo {year}
  {2017}{\natexlab{b}})}\BibitemShut {NoStop}%
\bibitem [{\citenamefont {Zhang}\ \emph {et~al.}(2019)\citenamefont {Zhang},
  \citenamefont {Ma}, \citenamefont {Wang}, \citenamefont {Sun}, \citenamefont
  {Lei}, \citenamefont {Lei}, \citenamefont {Chen}, \citenamefont {Wang},
  \citenamefont {Chen},\ and\ \citenamefont {Zhang}}]{Zhang:2019}%
  \BibitemOpen
  \bibfield  {author} {\bibinfo {author} {\bibfnamefont {A.}~\bibnamefont
  {Zhang}}, \bibinfo {author} {\bibfnamefont {X.}~\bibnamefont {Ma}}, \bibinfo
  {author} {\bibfnamefont {Y.}~\bibnamefont {Wang}}, \bibinfo {author}
  {\bibfnamefont {S.}~\bibnamefont {Sun}}, \bibinfo {author} {\bibfnamefont
  {B.}~\bibnamefont {Lei}}, \bibinfo {author} {\bibfnamefont {H.}~\bibnamefont
  {Lei}}, \bibinfo {author} {\bibfnamefont {X.}~\bibnamefont {Chen}}, \bibinfo
  {author} {\bibfnamefont {X.}~\bibnamefont {Wang}}, \bibinfo {author}
  {\bibfnamefont {C.}~\bibnamefont {Chen}}, \ and\ \bibinfo {author}
  {\bibfnamefont {Q.}~\bibnamefont {Zhang}},\ }\href {\doibase
  10.1103/PhysRevB.100.060504} {\bibfield  {journal} {\bibinfo  {journal}
  {Phys. Rev. B}\ }\textbf {\bibinfo {volume} {100}},\ \bibinfo {pages}
  {060504(R)} (\bibinfo {year} {2019})}\BibitemShut {NoStop}%
\bibitem [{\citenamefont {Muschler}\ \emph {et~al.}(2009)\citenamefont
  {Muschler}, \citenamefont {Prestel}, \citenamefont {Hackl}, \citenamefont
  {Devereaux}, \citenamefont {Analytis}, \citenamefont {Chu},\ and\
  \citenamefont {Fisher}}]{Muschler:2009}%
  \BibitemOpen
  \bibfield  {author} {\bibinfo {author} {\bibfnamefont {B.}~\bibnamefont
  {Muschler}}, \bibinfo {author} {\bibfnamefont {W.}~\bibnamefont {Prestel}},
  \bibinfo {author} {\bibfnamefont {R.}~\bibnamefont {Hackl}}, \bibinfo
  {author} {\bibfnamefont {T.~P.}\ \bibnamefont {Devereaux}}, \bibinfo {author}
  {\bibfnamefont {J.~G.}\ \bibnamefont {Analytis}}, \bibinfo {author}
  {\bibfnamefont {J.~H.}\ \bibnamefont {Chu}}, \ and\ \bibinfo {author}
  {\bibfnamefont {I.~R.}\ \bibnamefont {Fisher}},\ }\href {\doibase
  10.1103/Physrevb.80.180510} {\bibfield  {journal} {\bibinfo  {journal} {Phys.
  Rev. B}\ }\textbf {\bibinfo {volume} {80}},\ \bibinfo {pages} {180510(R)}
  (\bibinfo {year} {2009})}\BibitemShut {NoStop}%
\bibitem [{\citenamefont {Jost}\ \emph {et~al.}(2018)\citenamefont {Jost},
  \citenamefont {Scholz}, \citenamefont {Zweck}, \citenamefont {Meier},
  \citenamefont {B\"ohmer}, \citenamefont {Canfield}, \citenamefont
  {Lazarevi\ifmmode~\acute{c}\else \'{c}\fi{}},\ and\ \citenamefont
  {Hackl}}]{Jost:2018}%
  \BibitemOpen
  \bibfield  {author} {\bibinfo {author} {\bibfnamefont {D.}~\bibnamefont
  {Jost}}, \bibinfo {author} {\bibfnamefont {J.-R.}\ \bibnamefont {Scholz}},
  \bibinfo {author} {\bibfnamefont {U.}~\bibnamefont {Zweck}}, \bibinfo
  {author} {\bibfnamefont {W.~R.}\ \bibnamefont {Meier}}, \bibinfo {author}
  {\bibfnamefont {A.~E.}\ \bibnamefont {B\"ohmer}}, \bibinfo {author}
  {\bibfnamefont {P.~C.}\ \bibnamefont {Canfield}}, \bibinfo {author}
  {\bibfnamefont {N.}~\bibnamefont {Lazarevi\ifmmode~\acute{c}\else
  \'{c}\fi{}}}, \ and\ \bibinfo {author} {\bibfnamefont {R.}~\bibnamefont
  {Hackl}},\ }\href {\doibase 10.1103/PhysRevB.98.020504} {\bibfield  {journal}
  {\bibinfo  {journal} {Phys. Rev. B}\ }\textbf {\bibinfo {volume} {98}},\
  \bibinfo {pages} {020504(R)} (\bibinfo {year} {2018})}\BibitemShut {NoStop}%
\bibitem [{\citenamefont {B\"ohm}\ \emph {et~al.}(2018)\citenamefont {B\"ohm},
  \citenamefont {Kretzschmar}, \citenamefont {Baum}, \citenamefont {Rehm},
  \citenamefont {Jost}, \citenamefont {Ahangharnejhad}, \citenamefont
  {Thomale}, \citenamefont {Platt}, \citenamefont {Maier}, \citenamefont
  {Hanke}, \citenamefont {Moritz}, \citenamefont {Devereaux}, \citenamefont
  {Scalapino}, \citenamefont {Maiti}, \citenamefont {Hirschfeld}, \citenamefont
  {Adelmann}, \citenamefont {Wolf}, \citenamefont {Wen},\ and\ \citenamefont
  {Hackl}}]{Bohm:2018}%
  \BibitemOpen
  \bibfield  {author} {\bibinfo {author} {\bibfnamefont {T.}~\bibnamefont
  {B\"ohm}}, \bibinfo {author} {\bibfnamefont {F.}~\bibnamefont {Kretzschmar}},
  \bibinfo {author} {\bibfnamefont {A.}~\bibnamefont {Baum}}, \bibinfo {author}
  {\bibfnamefont {M.}~\bibnamefont {Rehm}}, \bibinfo {author} {\bibfnamefont
  {D.}~\bibnamefont {Jost}}, \bibinfo {author} {\bibfnamefont {R.~H.}\
  \bibnamefont {Ahangharnejhad}}, \bibinfo {author} {\bibfnamefont
  {R.}~\bibnamefont {Thomale}}, \bibinfo {author} {\bibfnamefont
  {C.}~\bibnamefont {Platt}}, \bibinfo {author} {\bibfnamefont {T.~A.}\
  \bibnamefont {Maier}}, \bibinfo {author} {\bibfnamefont {W.}~\bibnamefont
  {Hanke}}, \bibinfo {author} {\bibfnamefont {B.}~\bibnamefont {Moritz}},
  \bibinfo {author} {\bibfnamefont {T.~P.}\ \bibnamefont {Devereaux}}, \bibinfo
  {author} {\bibfnamefont {D.~J.}\ \bibnamefont {Scalapino}}, \bibinfo {author}
  {\bibfnamefont {S.}~\bibnamefont {Maiti}}, \bibinfo {author} {\bibfnamefont
  {P.~J.}\ \bibnamefont {Hirschfeld}}, \bibinfo {author} {\bibfnamefont
  {P.}~\bibnamefont {Adelmann}}, \bibinfo {author} {\bibfnamefont
  {T.}~\bibnamefont {Wolf}}, \bibinfo {author} {\bibfnamefont {H.~H.}\
  \bibnamefont {Wen}}, \ and\ \bibinfo {author} {\bibfnamefont
  {R.}~\bibnamefont {Hackl}},\ }\href {\doibase 10.1038/s41535-018-0118-z}
  {\bibfield  {journal} {\bibinfo  {journal} {Npj Quantum Mater.}\ }\textbf
  {\bibinfo {volume} {3}},\ \bibinfo {pages} {48} (\bibinfo {year}
  {2018})}\BibitemShut {NoStop}%
\bibitem [{\citenamefont {Devereaux}\ and\ \citenamefont
  {Einzel}(1995)}]{Devereaux:1995}%
  \BibitemOpen
  \bibfield  {author} {\bibinfo {author} {\bibfnamefont {T.~P.}\ \bibnamefont
  {Devereaux}}\ and\ \bibinfo {author} {\bibfnamefont {D.}~\bibnamefont
  {Einzel}},\ }\href {\doibase 10.1103/PhysRevB.51.16336} {\bibfield  {journal}
  {\bibinfo  {journal} {Phys. Rev. B}\ }\textbf {\bibinfo {volume} {51}},\
  \bibinfo {pages} {16336} (\bibinfo {year} {1995})}\BibitemShut {NoStop}%
\bibitem [{\citenamefont {Monien}\ and\ \citenamefont
  {Zawadowski}(1990)}]{Monien:1990}%
  \BibitemOpen
  \bibfield  {author} {\bibinfo {author} {\bibfnamefont {H.}~\bibnamefont
  {Monien}}\ and\ \bibinfo {author} {\bibfnamefont {A.}~\bibnamefont
  {Zawadowski}},\ }\href {\doibase 10.1103/PhysRevB.41.8798} {\bibfield
  {journal} {\bibinfo  {journal} {Phys. Rev. B}\ }\textbf {\bibinfo {volume}
  {41}},\ \bibinfo {pages} {8798} (\bibinfo {year} {1990})}\BibitemShut
  {NoStop}%
\bibitem [{\citenamefont {Suhl}\ \emph {et~al.}(1959)\citenamefont {Suhl},
  \citenamefont {Matthias},\ and\ \citenamefont {Walker}}]{Suhl:1959}%
  \BibitemOpen
  \bibfield  {author} {\bibinfo {author} {\bibfnamefont {H.}~\bibnamefont
  {Suhl}}, \bibinfo {author} {\bibfnamefont {B.~T.}\ \bibnamefont {Matthias}},
  \ and\ \bibinfo {author} {\bibfnamefont {L.~R.}\ \bibnamefont {Walker}},\
  }\href {\doibase 10.1103/PhysRevLett.3.552} {\bibfield  {journal} {\bibinfo
  {journal} {Phys. Rev. Lett.}\ }\textbf {\bibinfo {volume} {3}},\ \bibinfo
  {pages} {552} (\bibinfo {year} {1959})}\BibitemShut {NoStop}%
\bibitem [{\citenamefont {Khodas}\ and\ \citenamefont
  {Chubukov}(2012)}]{Khodas:2012}%
  \BibitemOpen
  \bibfield  {author} {\bibinfo {author} {\bibfnamefont {M.}~\bibnamefont
  {Khodas}}\ and\ \bibinfo {author} {\bibfnamefont {A.~V.}\ \bibnamefont
  {Chubukov}},\ }\href {\doibase 10.1103/PhysRevLett.108.247003} {\bibfield
  {journal} {\bibinfo  {journal} {Phys. Rev. Lett.}\ }\textbf {\bibinfo
  {volume} {108}},\ \bibinfo {pages} {247003} (\bibinfo {year}
  {2012})}\BibitemShut {NoStop}%
\bibitem [{\citenamefont {Thorsm{\o}lle}\ \emph {et~al.}(2016)\citenamefont
  {Thorsm{\o}lle}, \citenamefont {Khodas}, \citenamefont {Yin}, \citenamefont
  {Zhang}, \citenamefont {Carr}, \citenamefont {Dai},\ and\ \citenamefont
  {Blumberg}}]{Thorsmolle:2016}%
  \BibitemOpen
  \bibfield  {author} {\bibinfo {author} {\bibfnamefont {V.~K.}\ \bibnamefont
  {Thorsm{\o}lle}}, \bibinfo {author} {\bibfnamefont {M.}~\bibnamefont
  {Khodas}}, \bibinfo {author} {\bibfnamefont {Z.~P.}\ \bibnamefont {Yin}},
  \bibinfo {author} {\bibfnamefont {C.~L.}\ \bibnamefont {Zhang}}, \bibinfo
  {author} {\bibfnamefont {S.~V.}\ \bibnamefont {Carr}}, \bibinfo {author}
  {\bibfnamefont {P.~C.}\ \bibnamefont {Dai}}, \ and\ \bibinfo {author}
  {\bibfnamefont {G.}~\bibnamefont {Blumberg}},\ }\href {\doibase
  10.1103/Physrevb.93.054515} {\bibfield  {journal} {\bibinfo  {journal} {Phys.
  Rev. B}\ }\textbf {\bibinfo {volume} {93}},\ \bibinfo {pages} {054515}
  (\bibinfo {year} {2016})}\BibitemShut {NoStop}%
\bibitem [{\citenamefont {Gallais}\ \emph {et~al.}(2016)\citenamefont
  {Gallais}, \citenamefont {Paul}, \citenamefont {Chauvi\`ere},\ and\
  \citenamefont {Schmalian}}]{Gallais:2016}%
  \BibitemOpen
  \bibfield  {author} {\bibinfo {author} {\bibfnamefont {Y.}~\bibnamefont
  {Gallais}}, \bibinfo {author} {\bibfnamefont {I.}~\bibnamefont {Paul}},
  \bibinfo {author} {\bibfnamefont {L.}~\bibnamefont {Chauvi\`ere}}, \ and\
  \bibinfo {author} {\bibfnamefont {J.}~\bibnamefont {Schmalian}},\ }\href
  {\doibase 10.1103/PhysRevLett.116.017001} {\bibfield  {journal} {\bibinfo
  {journal} {Phys. Rev. Lett.}\ }\textbf {\bibinfo {volume} {116}},\ \bibinfo
  {pages} {017001} (\bibinfo {year} {2016})}\BibitemShut {NoStop}%
\bibitem [{\citenamefont {Kang}\ and\ \citenamefont
  {Fernandes}(2016)}]{Kang:2016}%
  \BibitemOpen
  \bibfield  {author} {\bibinfo {author} {\bibfnamefont {J.}~\bibnamefont
  {Kang}}\ and\ \bibinfo {author} {\bibfnamefont {R.~M.}\ \bibnamefont
  {Fernandes}},\ }\href {\doibase 10.1103/PhysRevLett.117.217003} {\bibfield
  {journal} {\bibinfo  {journal} {Phys. Rev. Lett.}\ }\textbf {\bibinfo
  {volume} {117}},\ \bibinfo {pages} {217003} (\bibinfo {year}
  {2016})}\BibitemShut {NoStop}%
\bibitem [{\citenamefont {Khodas}\ \emph {et~al.}(2014)\citenamefont {Khodas},
  \citenamefont {Chubukov},\ and\ \citenamefont {Blumberg}}]{Khodas:2014}%
  \BibitemOpen
  \bibfield  {author} {\bibinfo {author} {\bibfnamefont {M.}~\bibnamefont
  {Khodas}}, \bibinfo {author} {\bibfnamefont {A.~V.}\ \bibnamefont
  {Chubukov}}, \ and\ \bibinfo {author} {\bibfnamefont {G.}~\bibnamefont
  {Blumberg}},\ }\href {\doibase 10.1103/PhysRevB.89.245134} {\bibfield
  {journal} {\bibinfo  {journal} {Phys. Rev. B}\ }\textbf {\bibinfo {volume}
  {89}},\ \bibinfo {pages} {245134} (\bibinfo {year} {2014})}\BibitemShut
  {NoStop}%
\bibitem [{\citenamefont {Cea}\ and\ \citenamefont
  {Benfatto}(2016)}]{Cea:2016}%
  \BibitemOpen
  \bibfield  {author} {\bibinfo {author} {\bibfnamefont {T.}~\bibnamefont
  {Cea}}\ and\ \bibinfo {author} {\bibfnamefont {L.}~\bibnamefont {Benfatto}},\
  }\href {\doibase 10.1103/Physrevb.94.064512} {\bibfield  {journal} {\bibinfo
  {journal} {Phys. Rev. B}\ }\textbf {\bibinfo {volume} {94}},\ \bibinfo
  {pages} {064512} (\bibinfo {year} {2016})}\BibitemShut {NoStop}%
\bibitem [{\citenamefont {Huang}\ \emph {et~al.}(2018)\citenamefont {Huang},
  \citenamefont {Sigrist},\ and\ \citenamefont {Weng}}]{Huang:2018}%
  \BibitemOpen
  \bibfield  {author} {\bibinfo {author} {\bibfnamefont {W.}~\bibnamefont
  {Huang}}, \bibinfo {author} {\bibfnamefont {M.}~\bibnamefont {Sigrist}}, \
  and\ \bibinfo {author} {\bibfnamefont {Z.~Y.}\ \bibnamefont {Weng}},\ }\href
  {\doibase 10.1103/Physrevb.97.144507} {\bibfield  {journal} {\bibinfo
  {journal} {Phys. Rev. B}\ }\textbf {\bibinfo {volume} {97}},\ \bibinfo
  {pages} {144507} (\bibinfo {year} {2018})}\BibitemShut {NoStop}%
\bibitem [{\citenamefont {Linscheid}\ \emph {et~al.}(2016)\citenamefont
  {Linscheid}, \citenamefont {Maiti}, \citenamefont {Wang}, \citenamefont
  {Johnston},\ and\ \citenamefont {Hirschfeld}}]{Linscheid:2016}%
  \BibitemOpen
  \bibfield  {author} {\bibinfo {author} {\bibfnamefont {A.}~\bibnamefont
  {Linscheid}}, \bibinfo {author} {\bibfnamefont {S.}~\bibnamefont {Maiti}},
  \bibinfo {author} {\bibfnamefont {Y.}~\bibnamefont {Wang}}, \bibinfo {author}
  {\bibfnamefont {S.}~\bibnamefont {Johnston}}, \ and\ \bibinfo {author}
  {\bibfnamefont {P.~J.}\ \bibnamefont {Hirschfeld}},\ }\href {\doibase
  10.1103/PhysRevLett.117.077003} {\bibfield  {journal} {\bibinfo  {journal}
  {Phys. Rev. Lett.}\ }\textbf {\bibinfo {volume} {117}},\ \bibinfo {pages}
  {077003} (\bibinfo {year} {2016})}\BibitemShut {NoStop}%
\bibitem [{\citenamefont {Mishra}\ \emph {et~al.}(2016)\citenamefont {Mishra},
  \citenamefont {Scalapino},\ and\ \citenamefont {Maier}}]{Mishra:2016}%
  \BibitemOpen
  \bibfield  {author} {\bibinfo {author} {\bibfnamefont {V.}~\bibnamefont
  {Mishra}}, \bibinfo {author} {\bibfnamefont {D.~J.}\ \bibnamefont
  {Scalapino}}, \ and\ \bibinfo {author} {\bibfnamefont {T.~A.}\ \bibnamefont
  {Maier}},\ }\href {\doibase 10.1038/srep32078} {\bibfield  {journal}
  {\bibinfo  {journal} {Sci. Rep.}\ }\textbf {\bibinfo {volume} {6}},\ \bibinfo
  {pages} {32078} (\bibinfo {year} {2016})}\BibitemShut {NoStop}%
\bibitem [{\citenamefont {Shim}\ and\ \citenamefont {Duffy}(2002)}]{Shim:2002}%
  \BibitemOpen
  \bibfield  {author} {\bibinfo {author} {\bibfnamefont {S.~H.}\ \bibnamefont
  {Shim}}\ and\ \bibinfo {author} {\bibfnamefont {T.~S.}\ \bibnamefont
  {Duffy}},\ }\href {\doibase https://doi.org/10.2138/am-2002-2-314} {\bibfield
   {journal} {\bibinfo  {journal} {American Mineralogist}\ }\textbf {\bibinfo
  {volume} {87}},\ \bibinfo {pages} {318} (\bibinfo {year} {2002})}\BibitemShut
  {NoStop}%
\bibitem [{\citenamefont {Sathe}\ and\ \citenamefont
  {Dubey}(2007)}]{Sathe:2007}%
  \BibitemOpen
  \bibfield  {author} {\bibinfo {author} {\bibfnamefont {V.~G.}\ \bibnamefont
  {Sathe}}\ and\ \bibinfo {author} {\bibfnamefont {A.}~\bibnamefont {Dubey}},\
  }\href {\doibase 10.1088/0953-8984/19/38/382201} {\bibfield  {journal}
  {\bibinfo  {journal} {J. Phys.: Condens. Matter}\ }\textbf {\bibinfo {volume}
  {19}},\ \bibinfo {pages} {382201} (\bibinfo {year} {2007})}\BibitemShut
  {NoStop}%
\bibitem [{\citenamefont {Kretzschmar}\ \emph {et~al.}(2016)\citenamefont
  {Kretzschmar}, \citenamefont {Böhm}, \citenamefont {Karahasanović},
  \citenamefont {Muschler}, \citenamefont {Baum}, \citenamefont {Jost},
  \citenamefont {Schmalian}, \citenamefont {Caprara}, \citenamefont {Grilli},
  \citenamefont {Di~Castro}, \citenamefont {Analytis}, \citenamefont {Chu},
  \citenamefont {Fisher},\ and\ \citenamefont {Hackl}}]{Kretzschmar:20162}%
  \BibitemOpen
  \bibfield  {author} {\bibinfo {author} {\bibfnamefont {F.}~\bibnamefont
  {Kretzschmar}}, \bibinfo {author} {\bibfnamefont {T.}~\bibnamefont {Böhm}},
  \bibinfo {author} {\bibfnamefont {U.}~\bibnamefont {Karahasanović}},
  \bibinfo {author} {\bibfnamefont {B.}~\bibnamefont {Muschler}}, \bibinfo
  {author} {\bibfnamefont {A.}~\bibnamefont {Baum}}, \bibinfo {author}
  {\bibfnamefont {D.}~\bibnamefont {Jost}}, \bibinfo {author} {\bibfnamefont
  {J.}~\bibnamefont {Schmalian}}, \bibinfo {author} {\bibfnamefont
  {S.}~\bibnamefont {Caprara}}, \bibinfo {author} {\bibfnamefont
  {M.}~\bibnamefont {Grilli}}, \bibinfo {author} {\bibfnamefont
  {C.}~\bibnamefont {Di~Castro}}, \bibinfo {author} {\bibfnamefont {J.~G.}\
  \bibnamefont {Analytis}}, \bibinfo {author} {\bibfnamefont {J.~H.}\
  \bibnamefont {Chu}}, \bibinfo {author} {\bibfnamefont {I.~R.}\ \bibnamefont
  {Fisher}}, \ and\ \bibinfo {author} {\bibfnamefont {R.}~\bibnamefont
  {Hackl}},\ }\href {\doibase 10.1038/nphys3634} {\bibfield  {journal}
  {\bibinfo  {journal} {Nat. Phys.}\ }\textbf {\bibinfo {volume} {12}},\
  \bibinfo {pages} {560} (\bibinfo {year} {2016})}\BibitemShut {NoStop}%
\bibitem [{\citenamefont {Lederer}\ \emph {et~al.}(2020)\citenamefont
  {Lederer}, \citenamefont {Jost}, \citenamefont {B\"ohm}, \citenamefont
  {Hackl}, \citenamefont {Berg},\ and\ \citenamefont
  {Kivelson}}]{Lederer:2020}%
  \BibitemOpen
  \bibfield  {author} {\bibinfo {author} {\bibfnamefont {S.}~\bibnamefont
  {Lederer}}, \bibinfo {author} {\bibfnamefont {D.}~\bibnamefont {Jost}},
  \bibinfo {author} {\bibfnamefont {T.}~\bibnamefont {B\"ohm}}, \bibinfo
  {author} {\bibfnamefont {R.}~\bibnamefont {Hackl}}, \bibinfo {author}
  {\bibfnamefont {E.}~\bibnamefont {Berg}}, \ and\ \bibinfo {author}
  {\bibfnamefont {S.~A.}\ \bibnamefont {Kivelson}},\ }\href {\doibase
  10.1080/14786435.2020.1768604} {\bibfield  {journal} {\bibinfo  {journal}
  {Phil. Mag.}\ }\textbf {\bibinfo {volume} {100}},\ \bibinfo {pages} {2477}
  (\bibinfo {year} {2020})}\BibitemShut {NoStop}%
\end{thebibliography}%

\end{document}